\documentclass[review,11pt,a4paper,twoside]{elsarticle}

\usepackage{hyperref}
\usepackage{a4wide}
\usepackage{amsfonts}
\usepackage{amsmath}
\usepackage{amssymb}
\usepackage{amsthm}
\usepackage{natbib}
\usepackage{algorithm}
\usepackage{algorithmic}
\usepackage{graphicx}
\usepackage{color}
\usepackage{textcomp}
\usepackage{gensymb}
\usepackage{booktabs}
\usepackage[T1]{fontenc}
\usepackage[utf8]{inputenc}
\usepackage[margin=10pt,font=small,labelsep=endash]{caption}

\usepackage{pgfplots}
\usepgfplotslibrary{statistics}
\pgfplotsset{compat=1.14}
\usepackage{tikz}
\pgfplotsset{every tick label/.append style={font=\footnotesize}}

\renewcommand{\vec}[1]{\ensuremath \mathbf{\boldsymbol{#1}}}

\newcommand{\abs}[1]{\ensuremath{\left\vert#1\right\vert}}

\newcommand{\zb}[1]{\ensuremath{\boldsymbol{#1}}}

\renewcommand{\d}{\, \mbox{d}}

\numberwithin{equation}{section}
\numberwithin{table}{section}
\numberwithin{figure}{section}

\usepackage{subfigure}

\newcounter{todocounter}
\newcommand{\todo}[2][noisnotdefined]{
 \marginpar{\fcolorbox{black}{yellow}{\footnotesize\textbf{todo}}
 \ifthenelse{\equal{#1}{noisnotdefined}}{}{\textcolor{black}{\newline\tiny #1}}}
 \textbf{\ifthenelse{\equal{#2}{.}}
   {\fcolorbox{orange}{white}{\textcolor{orange}{$\maltese$}}}{{\textcolor{orange}{#2}}}}
 \refstepcounter{todocounter}}

\newcommand{\changed}[1]{\textcolor{orange}{#1}}
\usepackage[normalem]{ulem}
\newcommand{\removed}[1]{\sout{#1}}

\journal{journal}

\begin{document}

\begin{frontmatter}

\title{Gazing at crystal balls: Electron backscatter diffraction pattern analysis and cross correlation on the sphere}

\author[TU]{Ralf Hielscher\corref{RH}}
\ead{ralf.hielscher@mathematik.tu-chemnitz.de}
\author[TU]{Felix Bartel}
\author[ICL]{Thomas Benjamin Britton}
\ead{b.britton@imperial.ac.uk}
\cortext[RH]{Corresponding author}
\address[TU]{Fakult\"at für Mathematik, Technische Universit\"at Chemnitz, 09107, Chemnitz, Germany}
\address[ICL]{Department of Materials, Imperial College London, London, SW7 2AZ, UK}

\begin{abstract}
We present spherical analysis of electron backscatter diffraction (EBSD) patterns with two new algorithms: (1) \changed{band localisation and band profile analysis using the spherical Radon transform}; (2) \changed{orientation determination using spherical cross correlation}. These new approaches are formally introduced and their accuracies are determined using dynamically simulated patterns. We demonstrate their utility with an experimental dataset obtained from ferritic iron. Our results indicate that the analysis of EBSD patterns on the surface of the sphere provides an \changed{elegant method of revealing information} from these rich sources of crystallographic data.
\end{abstract}


\end{frontmatter}

\section*{Highlights}
\label{sec:Highlights}
\begin{enumerate}
\item \changed{We present a method to approximate Kikuchi patterns by
    spherical functions.}
\item \changed{We utilize the spherical Radon transformation to localise
    Kikuchi bands and to analyse their profile.}
\item We develop a spherical cross correlation method for orientation
  determination from Kikuchi patterns.
\item All methods are speed optimised using fast Fourier algorithms on the
  sphere and the orientation space.
\end{enumerate}


\section{Introduction}
\label{sec:introduction}

Electron backscatter diffraction (EBSD) is a popular microscopy technique used to reveal crystallographic information about materials. Automated, quantitative, robust, and precise interpretation of each electron backscatter pattern (EBSP) (aka Kikuchi pattern) has long been a major advantage of the technique \cite{Wilkinson2012}, especially when compared to other methods, such as transmission electron microscopy (TEM) imaging and until recently \cite{Ruach2010} S(canning)TEM based mapping. To advance the EBSD technique further, it is advantageous to improve the quality of the information captured and simultaneously improve how we interpret each pattern. The latter motivates our present work, where we consider the patterns as images on the sphere and analyse them using the spherical Radon transform and spherical cross correlation. In particular, both algorithms may be utilized for automatic orientation determination and band shape analysis.

The majority of existing algorithms for the analysis of EBSP consider them in the gnomonic projection, i.e., as they are measured  at a flat 2D capturing screen \cite{Adams1993}. In this manuscript, we exploit the fact that the EBSP is generated from a point source and is, therefore, more properly rendered onto the surface of the sphere \cite{Day2008}. For band localisation this is advantageous as Kikuchi bands have parallel edges on the sphere, but hyperbolic edges when considered in the plane. For the cross correlation method, the advantage of the spherical representation originates from the fact that different orientations differ just by a rotation of the spherical Kikuchi pattern while their correspondence at a flat detector is more involved.



The EBSP is generated as electrons enter the sample, scatter, and dynamically diffract. For an introduction to conventional EBSD analysis, see the review article by Wilkinson and Britton \cite{Wilkinson2012}. In practice, diffracting electrons are captured using a flat screen inserted within the electron microscope chamber. The result of this dynamical diffraction process is the generation of an EBSP that contains bands of raised intensity which are called the ``Kikuchi bands''. The centre line of each band corresponds to a plane that contains the electron source point and is parallel to the diffracting crystal plane. The edges of the bands are two Kossel conic sections separated by \( 2\theta\). The dynamical diffraction process is explained in greater detail in the work of Winkelmann et al.~\cite{Winkelmann2007}. \changed{The corresponding software provides us with} high quality simulations that contain significant crystallographic information, such as the intensity profile near a zone axes. These simulated patterns more accurately reproduce the intensity distributions found within experimentally captured patterns, as compared to simple kinematic models. This development has spurred an interest in using these patterns for direct orientation determination by \changed{pattern matching techniques} \cite{Chen2015}.

The Hough transform has been used to render the bands within the EBSP as points within a
transformed space for easy localisation using a computer \cite{Lassen1996}. In these conventional algorithms, \changed{it is assumed} that the bands within the EBSP are near parallel. This renders localisation of the bands into the computationally simpler challenge of finding peaks of high intensity within a sparsely populated space. Unfortunately, within the gnomonic projection the edges of the bands are not parallel. Additionally, the Hough transform of the bands produces butterfly artefacts which makes precise and robust localisation of the bands challenging. However, if the bands are presented as rings on the surface of the sphere \cite{Day2008} there is potential to integrate intensity profiles more precisely. This is advantageous for geometries where there may be divergence of the bands (e.g.~low voltage or where the pattern centre is less central).

To advance our analysis further, peak localisation and indexing may not be needed if we can efficiently directly compare and match the intensity distributions found within a high quality simulation against our experimental pattern. This can be performed with cross correlation (i.e.~finding a peak in the associated cross correlation function), which underpins template matching based EBSD analysis, including the ``dictionary indexing'' method \cite{Chen2015} and template matching approaches \cite{Foden2018,Wilkinson2018}.

Existing cross correlation methods \cite{WilkinsonMeadenDingley2006,Chen2015,Foden2018,Wilkinson2018} are performed within the gnomonic projection of the detector.  Hereby, each measured Kikuchi pattern $P=(P_{ij})$ is compared with a reference pattern $S(\zb O)=(S_{ij}(\zb O))$ according to a test orientation $\zb O$. The fit between both images is commonly measured by their correlation
\begin{equation}
  \label{eq:pm}
  C(P,S(\zb O)) = \sum_{i,j} P_{ij} S_{ij}(\zb O)
\end{equation}
where the sum is over all pixels in the pattern.

For template matching, reference patterns are tested according to multiple orientations. Sampling of the orientations is performed with a desired angular resolution (sufficient to find a peak and related to the ultimate angular sensitivity). This is computationally very expensive as the above sum has to be computed for a sufficiently large number of reference patterns $S(\zb O_{m})$, $m=1,\ldots,M$ to have a good estimate of the true orientation of the measured Kikuchi pattern $P$. Recently, Foden et al.~\cite{Foden2018} have presented an alternative approach where a FFT-based cross correlation is combined with a subsequent orientation refinement step to interpolate between library patterns to provide a more computationally efficient method of template matching. However, that method still involves an expensive gnomonic based library search. 

In this work we address this efficiency problem and perform the matching directly on the surface of
the sphere. Therefore, we require only one spherical master pattern.
In this paradigm different orientations results in different rotations of the spherical master pattern. The central idea of this paper is to represent the correlation function between the experimental Kikuchi pattern and all rotated versions of the spherical master pattern as a spherical convolution which can be computed  using fast Fourier techniques.

In the case of plane images $P$ and $S$ it is well known \cite{Anuta1970} that the correlation image
\begin{equation*}
    C_{k,\ell} = \sum_{i,j} P_{ij} S_{i+k,j+\ell}
\end{equation*}
with respect to all shifts $k,\ell$ can be computed simultaneously using the fast Fourier transform $\mathcal F$. More precisely, we have
\begin{equation*}
  C = \mathcal F^{-1}(\mathcal F P \odot \mathcal F S)
\end{equation*}
where $\odot$ denotes the pointwise product. Such Fourier based algorithms have approximately square root the number of operations compared with direct algorithms.

The match between two spherical diffraction patterns can be measured through
spherical cross correlation resulting in a function on orientation space. The
position of the maximum peak of this function directly gives the desired misorientation of the
experimental pattern with respect to the master pattern. In order to speed up
the computation of the spherical cross correlation function we apply the same
Fourier trick as explained above. In short, we compute spherical Fourier
coefficients of the experimental and the master pattern, multiply them and
obtain a series representation of the cross correlation function with respect
to generalised spherical harmonics. Computation of the spherical Fourier
coefficients and evaluation of the generalised spherical harmonics is done
using the nonequispaced fast Fourier transform (NFFT) which is at the heart of the
MTEX toolbox used for crystallographic texture analysis. The NFFT builds upon significant
research generalising the FFT to non Euclidean domains, e.g.~to the sphere,
cf.\,\cite{Kunis2009,Kunis2003}, or the orientation space
cf.\,\cite{Potts2009} and to apply them to problems in quantitative texture
analysis,
cf.\,\cite{HielscherPotts2008,Hielscher2010,Mainprice2015a,Mainprice2015b}.
Although our algorithm\changed{s} are theoretically fast the running times of our implementations
are behind those of well established methods. The main reason for this is that our implementations
are not yet optimised to take advantage of crystal symmetries, computing on the graphics card or any other kind of parallelisation.
On the other hand, this keeps our proof of concept code very simple and allows for easy customisation.

\section{Spherical diffraction patterns}
\label{Spherical}
The advantages of considering Kikuchi patterns as spherical functions have
been explained very nicely by Day \cite{Day2008}. As an illustrative example of a Kikuchi pattern we consider a high quality dynamical simulation of $\alpha$-Iron (BCC)
generated within DynamicS (Bruker Nano GmbH) and project it onto the surface of the sphere (Figure \ref{fig:quadrature}a). The
commercial program uses dynamical theory presented by Winkelmann et
al.~\cite{Winkelmann2007} to calculate the intensity of electrons in the
resultant diffraction pattern.

In the case of experimental patterns, the diffraction sphere is not completely
described as the detector does not subtend all diffraction angles (Figure \ref{fig:S2ApproxPattern}). The amount
of the sphere covered is described by the shape, size\changed{,} and detector distance.
For our algorithms the incomplete coverage causes two issues: (1) edge effects, which can be resolved by appropriate use of windowing functions; (2) incomplete Kikuchi bands which leads
to different peak intensities in the spherical Radon transform.
We will address these in more detail within Sec.~4.

\section{Harmonic approximation on the sphere}
\label{SphereFT}

\begin{figure}
  \centering
  \subfigure[simulated]{
    \includegraphics[width=0.23 \textwidth]{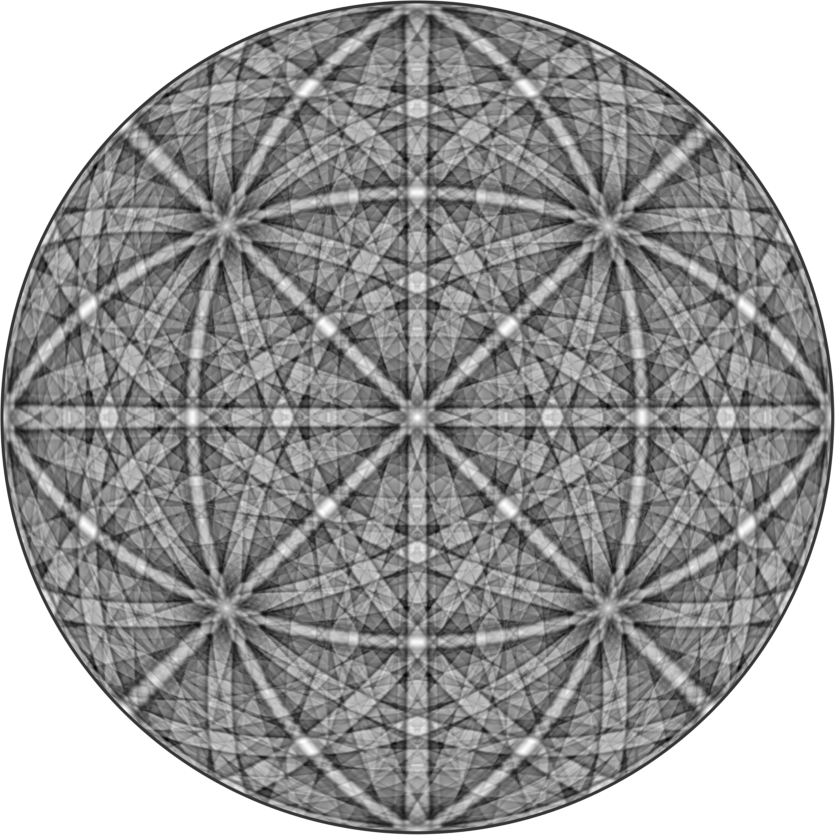}}
  \subfigure[$N=256$]{
    \includegraphics[width=0.23 \textwidth]{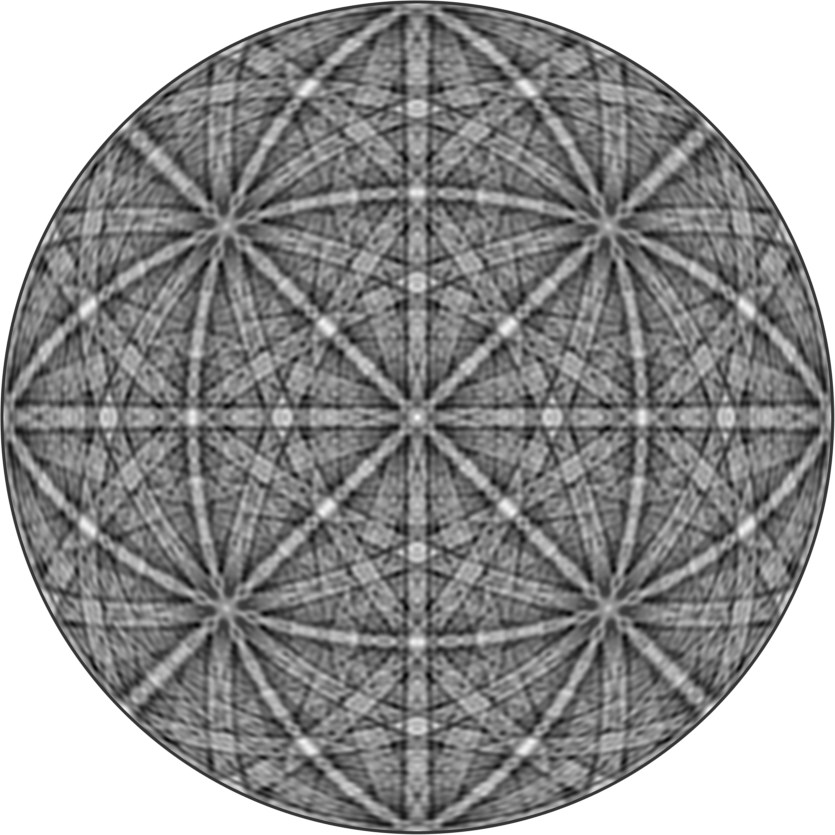}}
  \subfigure[$N=128$]{
    \includegraphics[width=0.23 \textwidth]{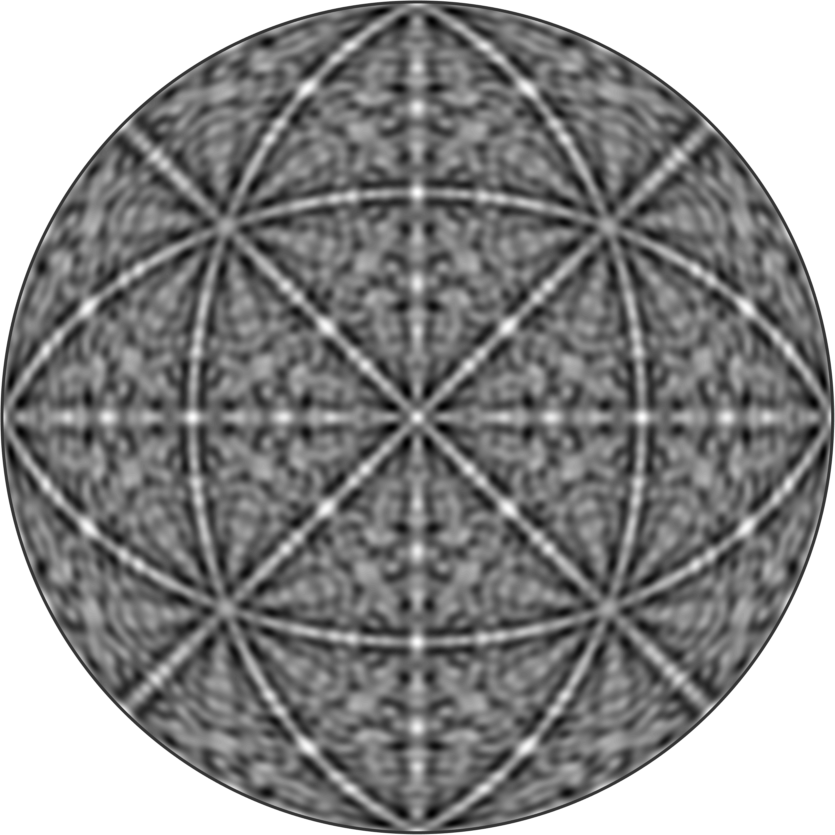}}
  \subfigure[$N=64$]{
    \includegraphics[width=0.23 \textwidth]{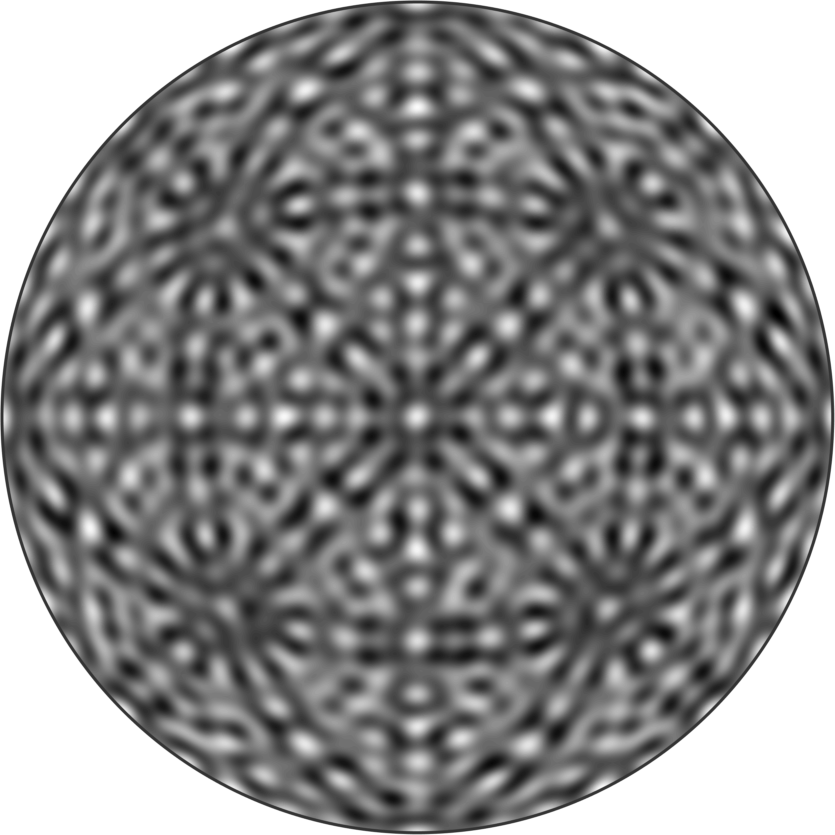}}

  \caption{Stereographic projection of the dynamically simulated Kikuchi pattern for iron (a) and its approximation by spherical harmonics with different harmonic degrees $N$.}
  \label{fig:quadrature}
\end{figure}

Simulated as well as experimental Kikuchi pattern\changed{s} can be interpreted as
diffraction intensities $f_j$ with respect to discrete diffraction directions $\vec \xi_{j}$
that can be computed from the position within the pattern by the inverse gnonomic projection.
For our algorithms, we are interested in approximating these intensities $f_j$ by a smooth spherical function
$f \colon \mathbb S^{2} \to \mathbb R$, represented by
a series expansion of the form
\begin{equation}
  \label{eq:S2Sum}
  f(\boldsymbol{\xi})
  = \sum_{n=0}^{N} \sum_{k=-n}^{n} \hat f(n,k) Y_{n}^{k}(\boldsymbol{\xi}),
\end{equation}
such that $f(\vec \xi_j)\approx f_j$. Hereby, $Y_{n}^{k}$ denotes the spherical harmonics  which replace the
exponential functions in the classical Fourier transform. \changed{Elegant introductions
into harmonic analysis on the sphere can be found in \cite{frgesc,Mic13}.} Similarly to the
classical case many properties of the function $f$ can be directly derived
from its Fourier coefficients $\hat f(n,k)$. If we consider $f$ as an image on
the sphere (see Figure \ref{fig:quadrature}a), many image operations, like convolution, rotation, or
differentiation, can be efficiently described in terms of the Fourier
coefficients .

There exist several methods for determining the coefficients $\hat f(n,k)$ in
\eqref{eq:S2Sum} from discrete diffraction intensities $f(\vec
\xi_{j})$. Three of those will be introduced shortly: quadrature,
interpolation, and approximation.

\subsection{Quadrature}
\label{sec:quadrature}

The quadrature based approach exploits the fact that the spherical harmonics
$Y_{n}^{k}$ form an orthonormal basis with respect to the inner product
\begin{equation*}
  \langle f_{1},f_{2} \rangle = \int_{\mathbb S^{2}} f_{1}(\vec \xi)
  \overline{f_{2}(\vec \xi)} \d{\vec \xi}.
\end{equation*}
As a consequence, the expansion coefficients $\hat f(n,k)$ satisfy
\begin{equation*}
  \hat f(n,k)
  = \int_{\mathbb S^{2}} f(\vec \xi) \overline{Y_{n}^{k}(\vec \xi)} \d{\vec \xi}.
\end{equation*}
Computing this integral numerically is called quadrature and leads to sums of the form
\begin{equation}
  \label{eq:quadrature}
  \hat f(n,k)
  \approx \sum_{j=1}^{J}  \omega_{j} f(\vec \xi_{j}) \overline{Y_{n}^{k}(\vec \xi_{j})},
\end{equation}
with the quadrature nodes $\vec \xi_{j} \in \mathbb S^{2}$ and quadrature
weights $\omega_{j} \in \mathbb R$, $j=1,\ldots,J$. The challenge is to find
those nodes and weights such that the approximation is as good as
possible. Good choices are discussed in \cite{Graf2010,Graf2013} and the
references therein.

Evaluating the sum \eqref{eq:quadrature} for $n=0,\ldots,N$ and
$k=-n,\ldots,n$ directly would require $N^2 \cdot J$ numerical
operations. Fortunately, this sum can be computed much faster using the
nonequispaced fast Fourier transform \cite{Keiner2009} requiring only
$N^2 \log N + J$ numerical operations.

The key parameter when approximating a spherical function by its harmonic
series expansion is the cut-off frequency $N$.  Figure \ref{fig:quadrature}
illustrates the effect of this cut-off frequency $N$ when approximating a
Kikuchi pattern. Our numerical experiments will show that a cut-off frequency of $N = 128$ provides
enough detail to \changed{enable band detection and orientation determination
  by cross correlation with reasonable precision for a typical pattern.}

The advantage of the quadrature based approach is its simplicity. This comes
at the cost that the function values of $f$ have to be known at the specific
quadrature nodes $\vec \xi_{j}$, which can be true for simulated patterns but
will not be true for experimental patterns.

\subsection{Interpolation and approximation}
\label{sec:interpolation}

If the function $f$ to be approximated is given at discrete points
$\vec \xi_{j}$, $j=1,\ldots,J$, i.e., $f(\vec \xi_{j}) = f_{j}$, for which no
quadrature rule is known we may compute the expansion coefficients
$\hat f(n,k)$ by solving the system of linear equations
\begin{equation} \label{eq:interpolation}
\sum_{n=0}^{N} \sum_{k=-n}^{n} \hat f(n,k) Y_{n}^{k}(\vec \xi_{j}) = f_{j}.
\end{equation}
It should be noted that this system of linear equations may become ill
conditioned, especially in the case that the number of interpolation nodes $J$
equals the number $(N+1)^{2}$ of coefficients $\hat f(n,k)$. It is therefore
recommended to consider the underdetermined or overdetermined problem and
solve it using the normal equation of first or second kind, respectively.

Interpolation corresponds to the underdetermined case where the system of
equations \eqref{eq:interpolation} has no unique solution.  To restore uniqueness
we search for coefficients solving
\eqref{eq:interpolation} and simultaneously minimising some functional
$\varphi(\vec{\hat f})$ which characterises the smoothness of $f$.
Common choices are Sobolev norms of order $s>0$,
\begin{equation*}
\varphi(\vec{\hat f}) = \sum_{n=0}^N\sum_{k=-n}^n (n+1)^s \left|\hat f(n,k)\right|^2.
\end{equation*}
The solution of this constrained minimisation problem can be found by solving the
corresponding normal equation of second
kind.  See also \cite{Keiner2007} for more details on the stability of
spherical interpolation.

In the case of experimentally measured data it can be easier and more stable
to solve an approximation problem instead of an interpolation problem,
i.e.\,we are in the overdetermined case and the system of equations
\eqref{eq:interpolation} must not have any solution.  We therefore look for
the coefficients $\hat f(n,k)$ which achieve the smallest error
\begin{equation*}
  F(\hat f)
  =  \sum_{j=1}^{J} \left(\sum_{n=0}^{N} \sum_{k=-n}^{n} \hat f(n,k)
    Y_{n}^{k}(\vec \xi_{j}) - f_{j}\right)^{2}
  + \lambda \sum_{n=0}^{N} \sum_{k=-n}^{n}  (n+1) \abs{\hat f(n,k)}^{2}
\end{equation*}
while decaying to zero quickly.  Here the first summand measures the fitting
of the approximation in the points $\vec\xi_j$ and the second summand is the
regularisation term that measures the smoothness of the function and punishes
noise. The weighting between these two terms is accomplished via the parameter
$\lambda$ which is often called regularisation parameter and has to be chosen
such that there is balance between these two contradicting terms.

\begin{figure}
  \centering
  \subfigure[\label{fig:patternRaw} Kikuchi pattern]{
    \includegraphics[width = 0.3\textwidth]{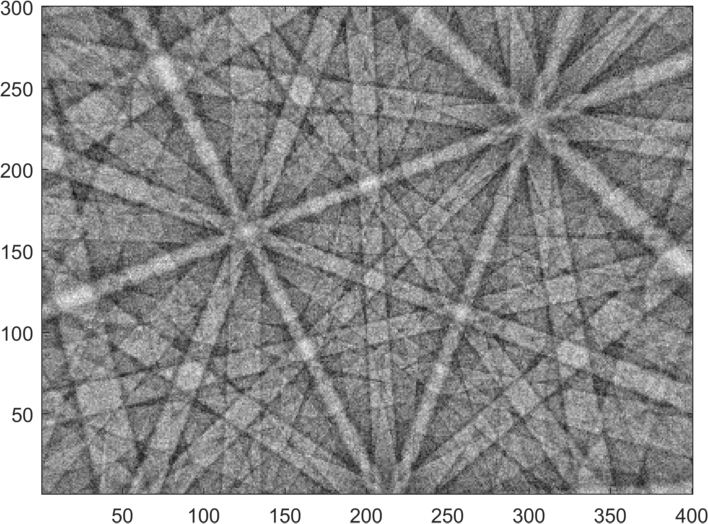}}

  \subfigure[\label{fig:patternQuat}quadrature]{\includegraphics[width =
    0.3\textwidth]{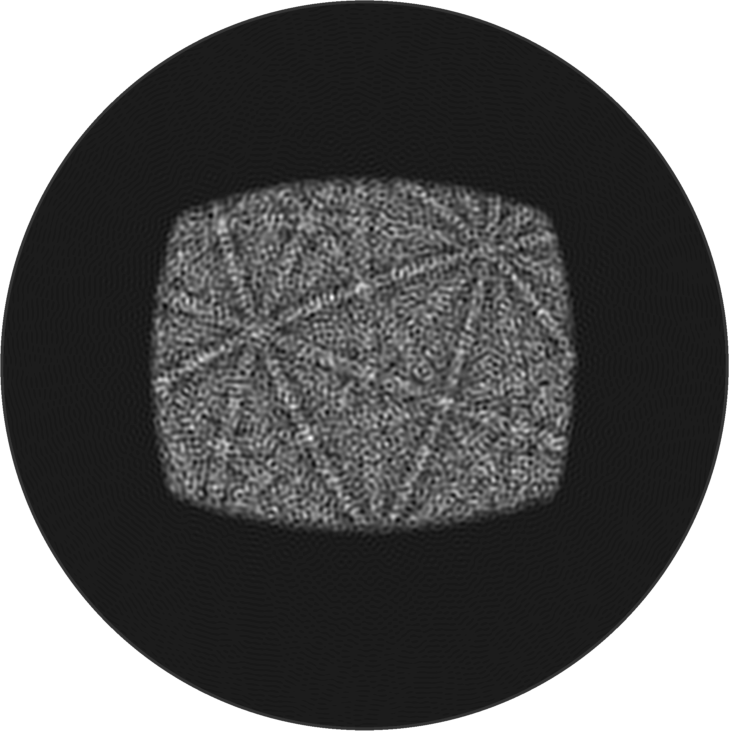}}
  \subfigure[\label{fig:patternApprox}approximation]{\includegraphics[width =
    0.3\textwidth]{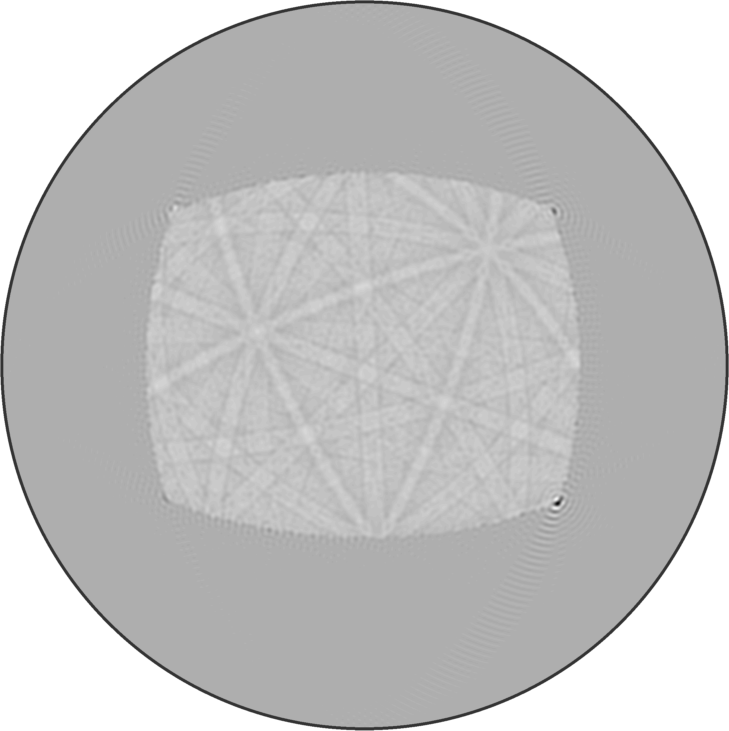}}
  \subfigure[\label{fig:patternApproxCutOff} with smooth cut-off function
  ]{\includegraphics[width = 0.3\textwidth]{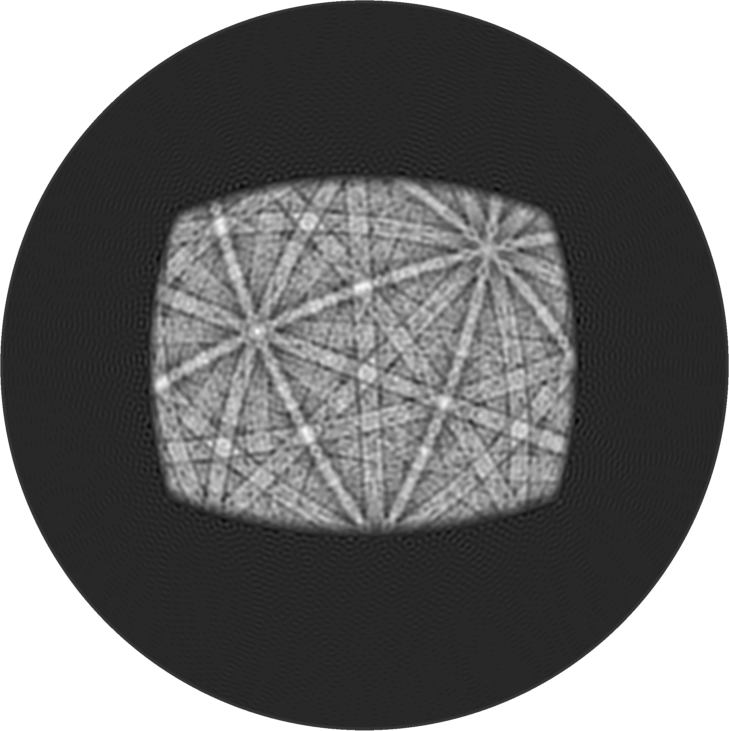}}

  \caption{Spherical approximations of a simulated Kikuchi pattern at a
    detector corrupted by noise.}
  \label{fig:S2ApproxPattern}
\end{figure}

As an example, Figure~\ref{fig:S2ApproxPattern} depicts an ``experimental''
Kikuchi pattern \ref{fig:patternRaw} together with a quadrature based
\ref{fig:patternQuat} and approximation based \ref{fig:patternApprox}
representation with respect to spherical harmonics. We observe that the
approximation based approach leads to severe artefacts close to the detector
boundaries. The reason is that harmonic functions are very bad in representing
functions with hard jumps. This problem can be significantly relaxed by
multiplying the data with a filter that generates a smooth decay from the values
inside the detector to zero outside the detector. The resulting harmonic
approximation is displayed in \ref{fig:patternApproxCutOff}.

\section{ \changed{Spherical Radon transform based band detection}}
\label{sec:spher-band-detect}

In conventional \changed{orientation determination from EBSD data} the Kikuchi
pattern is represented in the flat, gnomonic frame and summed up along all straight
lines resulting in the Radon (or Hough) transform. Since in the Radon transform
diffraction bands appear as local extrema they can be found by a peak
detection algorithm. A severe problem of this approach is that due to the
gnomonic projection bands, in the Kikuchi pattern, do not appear as straight
features but have hyperbolic shape. As a consequence the local extrema are
less sharp which negatively affects the accuracy and robustness of this
approach. An alternative band analysis method which correctly uses the fact
that the parallel bands on the surface of the sphere are well represented as
hyperbolic sections in the gnomonic frame is incorporated in the 3D Hough
transform \cite{Maurice2008}.

In this section we will make use of the fact that Kikuchi bands on the sphere
are centered around great circles with edges formed by small circles that can
be efficiently detected by a spherical Radon transform and its
generalisations. Once sufficiently many bands are located the orientation can
be determined by conventional indexing algorithms, e.g.~with
AstroEBSD~\cite{Britton2018}.

\subsection{The spherical Radon transform}
\label{sec:spher-appr-1}
The spherical Radon transform integrates a function on the sphere along all great circles, which is similar to how the ordinary Radon transform integrates an
image along all lines. Such a great circle $\mathcal C$ on the
sphere can be described as the set of all points $\vec \xi \in \mathbb S^{2}$
that are orthogonal to a given normal vector $\vec \eta \in \mathbb S^{2}$,
i.e.,
$\mathcal C(\vec \eta) = \{\vec \xi \in \mathbb S^{2} \mid \vec \xi \cdot \vec
\eta = 0\}$. Accordingly the spherical Radon transform
\begin{equation}
  \label{eq:DefRadonS2}
  g(\vec \eta) = \mathcal R f(\vec \eta)
  = \int_{\mathcal C(\vec \eta)}
  f(\vec \xi) \, \mathrm{d}(\vec \xi)
\end{equation}
of a spherical function $f \colon \mathbb S^{2} \to \mathbb R$ is again a
spherical function $g \colon \mathbb S^{2} \to \mathbb R$.

The crucial point is now, that the Fourier representation of $g$ can be
computed straight forward from the Fourier coefficients $\hat f(n,k)$ of $f$,
i.e., we have
\begin{equation}
  \label{eq:RadonS2}
  g(\boldsymbol{\eta})
  = \sum_{n=0}^{N} \sum_{k=-n}^{n} P_{n}(0) \hat f(n,k) Y_{n}^{k}(\boldsymbol{\eta}),
\end{equation}
where $P_{n}(0)$ are the Legendre polynomials evaluated in the point $0$. The
practical use of this formula is that for computing the Radon transform of a
spherical image, we do not need to average the pixel values along all great
circles but, instead, compute the Fourier coefficients of the spherical image,
multiply them with
\begin{equation*}
  P_{n}(0) =
  \begin{cases}
    (-1)^{n/2}\frac{(n-1)(n-3)\cdots 3 \cdot 1}{n(n-2)\cdots 4 \cdot 2}, & \quad n \text{ is even }\\
    0, & \quad n \text{ is odd}
  \end{cases}
\end{equation*}
and apply the spherical Fourier transform which gives us the spherical image
of the Radon transform. For an image of $m \times m$ pixels the later
algorithm using the nonequispaced fast spherical Fourier transform
\cite{Kunis2003} is about $m$ times faster. 

\definecolor{c1}{rgb}{ 0.1216, 0.4667, 0.7059}
\definecolor{c2}{rgb}{ 1.0000, 0.4980, 0.0549}
\definecolor{c3}{rgb}{ 0.1725, 0.6275, 0.1725}
\definecolor{c4}{rgb}{0.8392, 0.1529, 0.1569}
\definecolor{c5}{rgb}{0.5804, 0.4039, 0.7412}

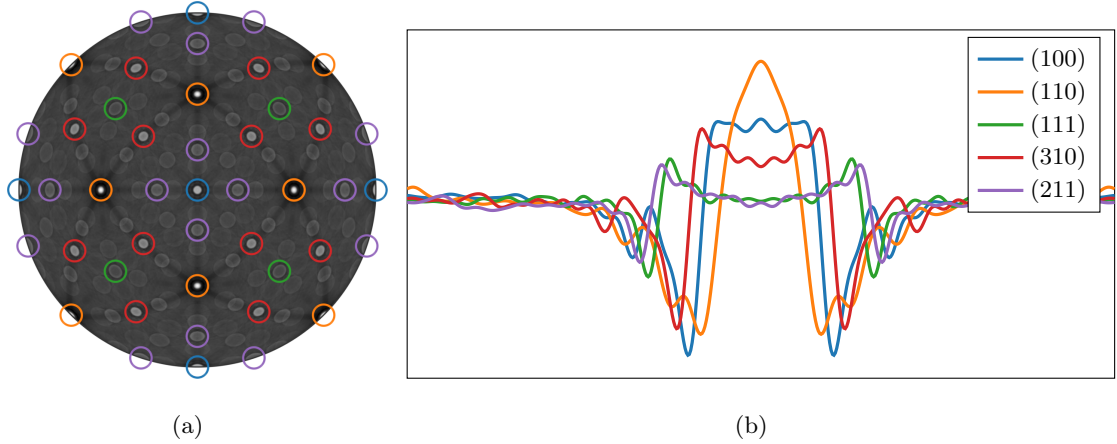
\begin{figure}
  \centering
  \subfigure[\label{fig:RadonMaster}]{
    \includegraphics[height = 5cm]{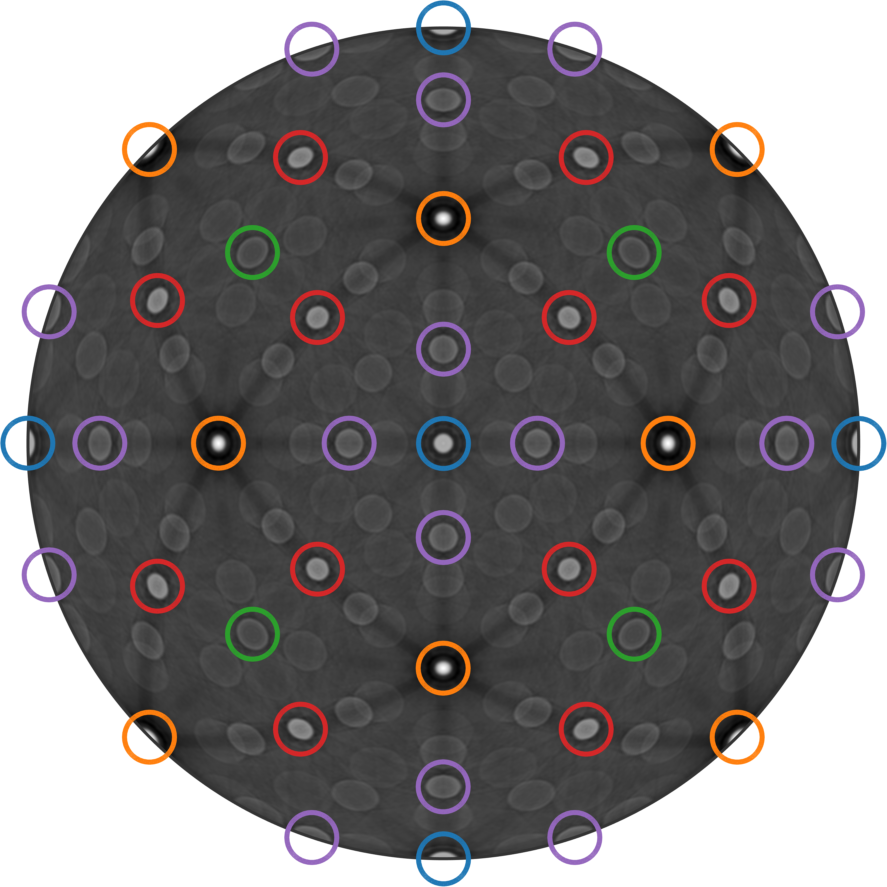}}
  \subfigure[]{    \begin{tikzpicture}
      \begin{axis}[width=0.7\textwidth,height=6.2cm,
        ymin=-0.25,ymax=0.25,xmin=80,xmax=100,enlargelimits=false,ticks=none,
      legend style={font=\footnotesize},
      xlabel style={font=\footnotesize},
      ]
      \addplot+[color=c1,mark=none, very thick] table[x index=0,y index=1]{./pic/profiles.txt};
      \addplot+[color=c2,mark=none, very thick] table[x index=0,y index=2]{./pic/profiles.txt};
      \addplot+[color=c3,mark=none, very thick] table[x index=0,y index=3]{./pic/profiles.txt};
      \addplot+[color=c4,mark=none, very thick] table[x index=0,y index=6]{./pic/profiles.txt};
      \addplot+[color=c5,mark=none, very thick] table[x index=0,y index=7]{./pic/profiles.txt};
      \legend{$(100)$, $(110)$,  $(111)$, $(310)$, $(211)$, $(321)$}

    \end{axis}
  \end{tikzpicture}}
  \caption{(a) Spherical Radon transform of the master pattern (b) band
    profiles corresponding to different lattice planes.}
  \label{fig:RadonS2}
\end{figure}

Fig.~\ref{fig:RadonS2}a shows the spherical Radon transform of the
dynamically simulated master pattern from Fig.~\ref{fig:quadrature}.
The circular features correspond to the bands in the Kikuchi pattern.

\subsection{Spherical convolution and band localisation}
\label{sec:integrals-over-small}

The brightness and sharpness of the Radon peaks is not uniform in
Fig.~\ref{fig:RadonS2}a due to the different shape of the bands corresponding
to the different lattice planes. To visualize and analyze the profile of the
band, corresponding to a plane with normal vector
$\vec \eta \in \mathbb S^{2}$, in more detail we integrate the spherical
diffraction pattern $f_{\text{sim}}$ in Fig.~\ref{fig:quadrature}c with
respect to all rotations $R_{\vec \eta}(\omega)$ about the plane normal
$\vec \eta$, i.e.,
\begin{equation*}
  \Phi_{\vec \eta}(\vec \xi)
  = \int_{0}^{2\pi} f_{\text{sim}}(R_{\vec \eta}(\omega) \vec \xi) \d{\omega}.
\end{equation*}
The resulting band profiles $\Phi_{\vec \eta}$ for the major bands 
are depicted in Fig.~\ref{fig:RadonS2}b.

Let us give a small site note on how those integrals can be computed
efficiently from the Fourier coefficients $\hat f_{\text{sim}}(n,k)$ of the
master pattern. In case the plane normal $\vec \eta$ coincides with the
$z$-axis the profile $\Phi_{\vec z}$ is given by the Legendre series
\begin{equation}
  \label{eq:4}
  \Phi_{\vec z}(\vec \xi)
  = \sum_{n=0}^{N} \hat f_{\text{sim}}(n,0) P_{n}(\vec \xi \cdot \vec z).
\end{equation}
In the general case of an arbitrary plane normal $\vec \eta$, it is sufficient to rotate
$f_{\text{sim}}$ such that the plane normal aligns with the $z$-axis and to
proceed as above.

We may use our knowledge of these band profiles to identify specific bands
within the experimental pattern using a spherical convolution
\begin{equation}
  \label{eq:S2Conv}
  f \star \Phi(\vec \eta)
  = \int_{\mathbb S^{2}} f(\vec \xi) \Phi(\vec \xi \cdot \vec \eta) \d{\vec \xi}.
\end{equation}
of the pattern $f$ with a specific band profile $\Phi$. The spherical
convolution with a band profile can be seen as a generalisation of the
spherical Radon transform. Indeed, choosing as the band profile
$\Phi = \delta$ the delta distribution the spherical convolution
$f \star \Phi = \mathcal R f$ coincides with the spherical Radon transform. On
the other hand, it may also be interpreted as a generalisation of the
butterfly mask \cite{Lassen1996} and the top hat filter \cite{Pinard2011} used
in conventional Radon/Hough based EBSD.

\begin{figure}
  \centering
    \subfigure[\label{fig:conv211}band (211)]{
    \begin{minipage}{0.22\linewidth}
    \includegraphics[width=\textwidth]{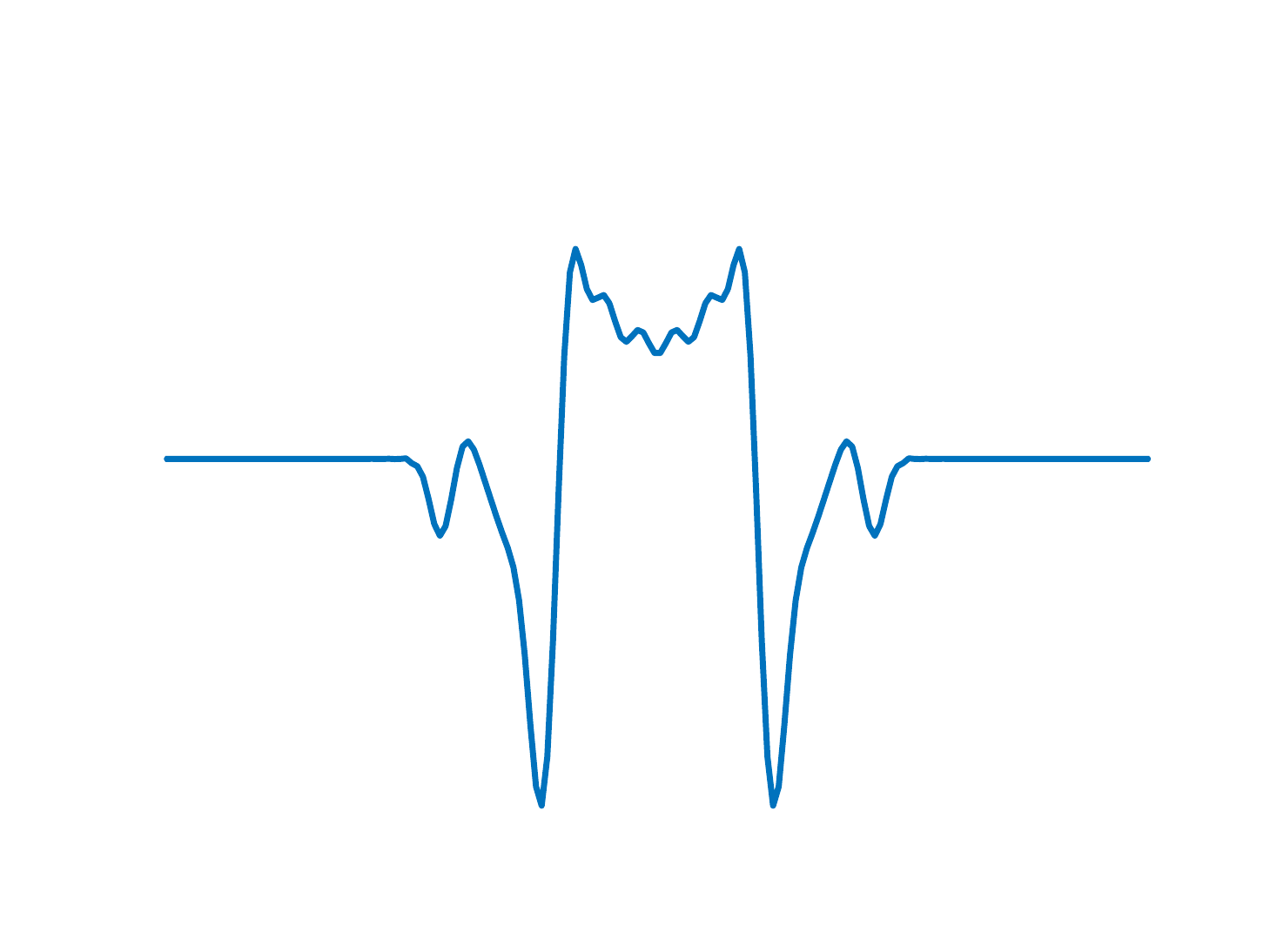}\\
    \includegraphics[width=\textwidth]{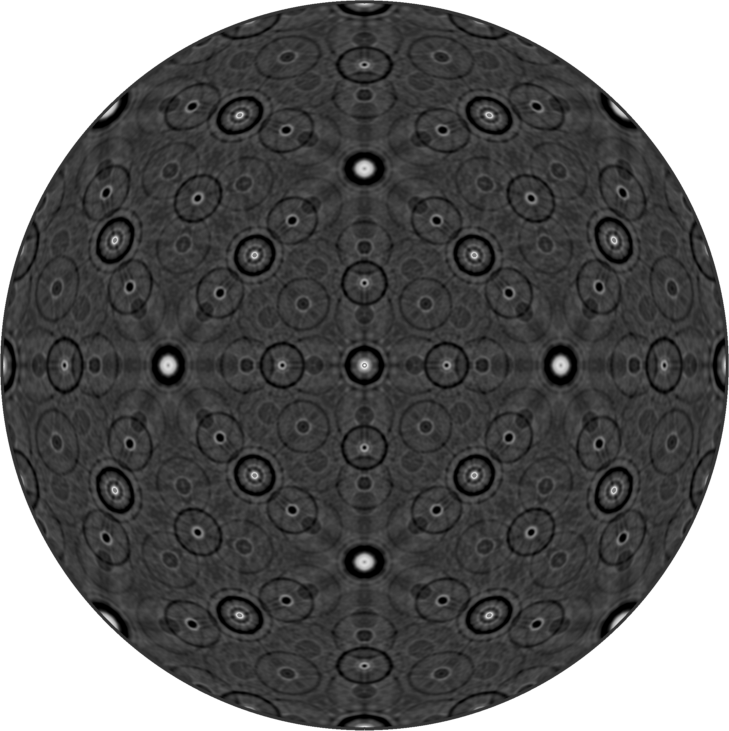}
    \end{minipage}
  }
    \subfigure[\label{fig:conv310}band (310)]{
    \begin{minipage}{0.22\linewidth}
    \includegraphics[width=\textwidth]{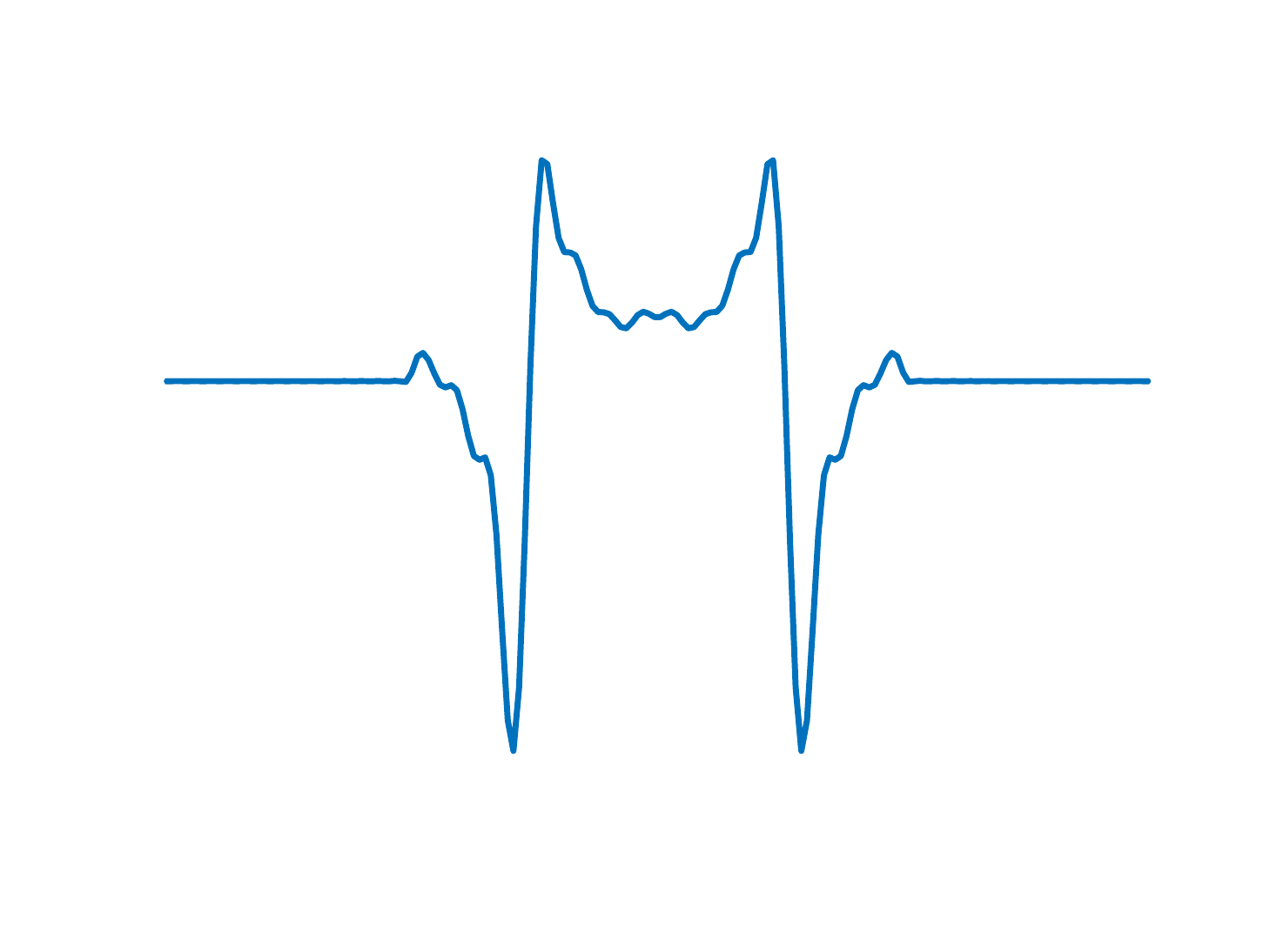}\\
    \includegraphics[width=\textwidth]{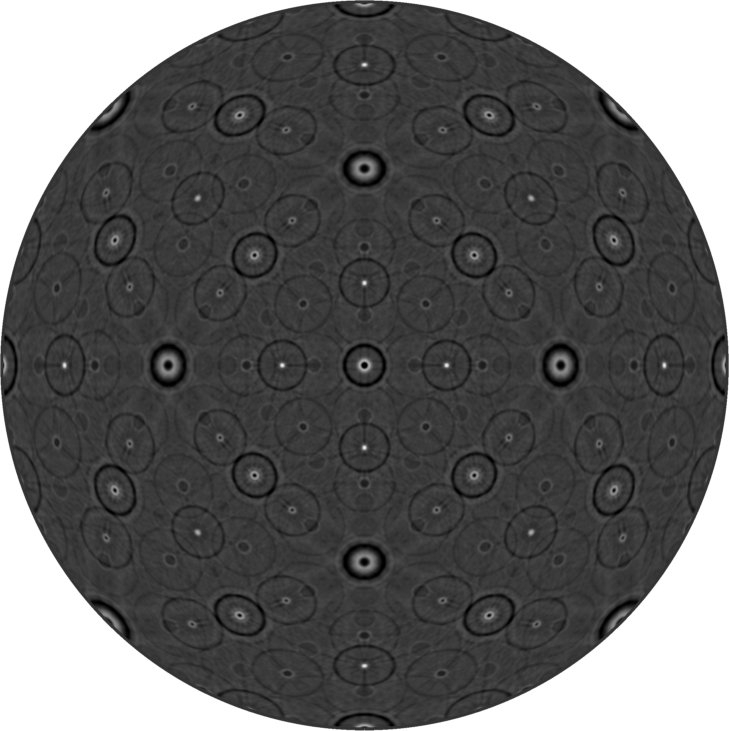}
    \end{minipage}
  }
  \subfigure[\label{fig:conv1} Gaussian]{
    \begin{minipage}{0.22\linewidth}
    \includegraphics[width=\textwidth]{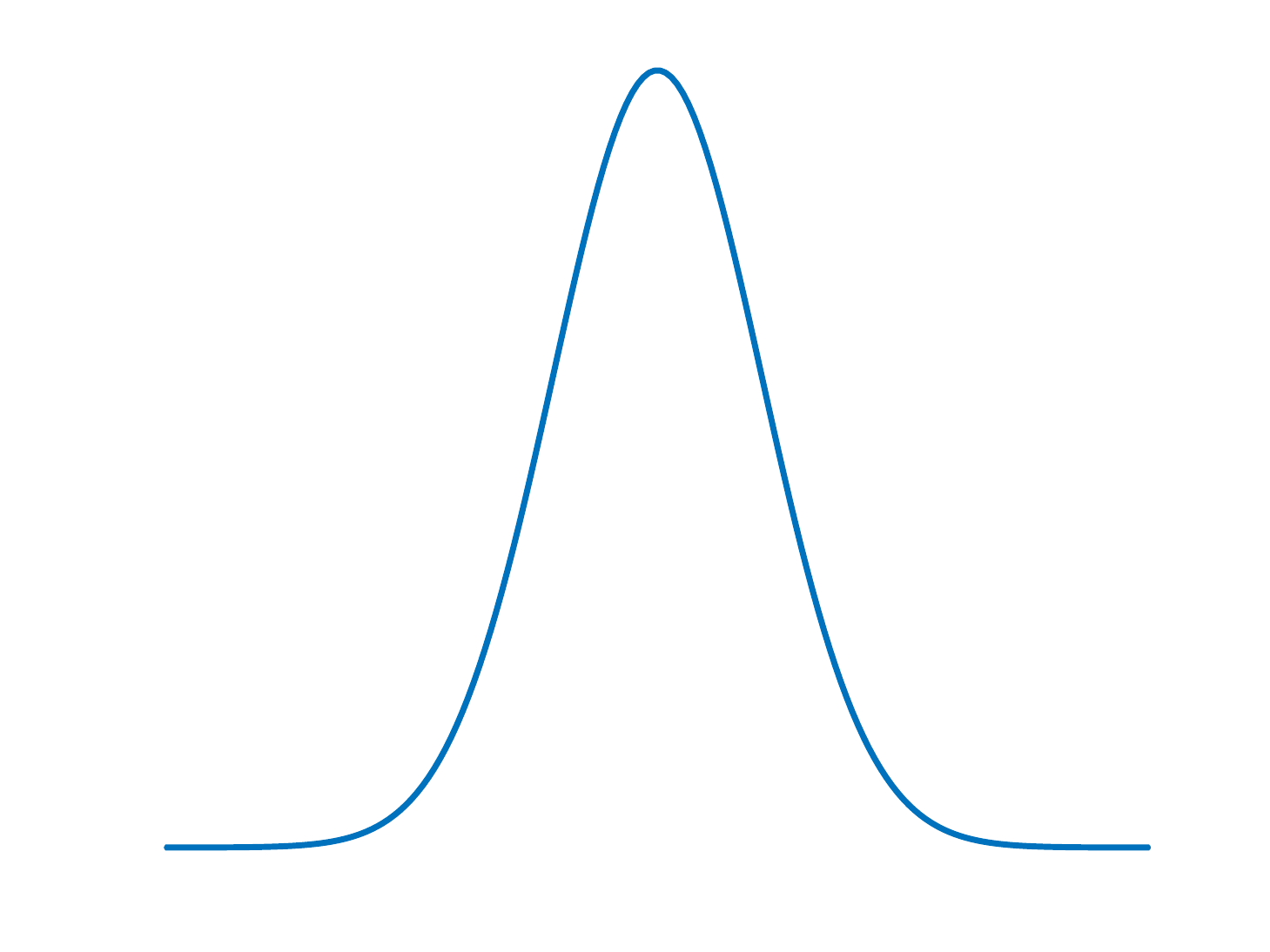}\\
    \includegraphics[width=\textwidth]{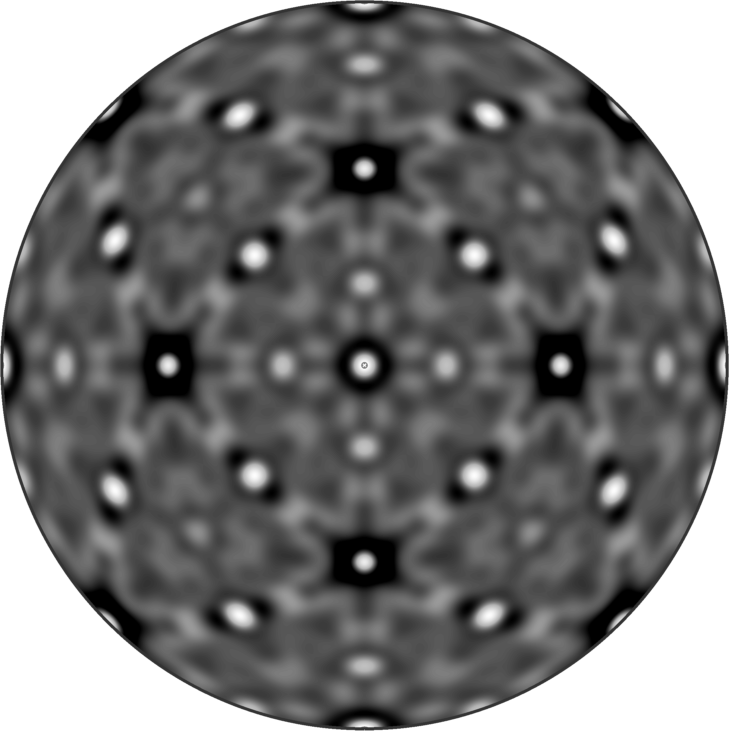}
    \end{minipage}
  }
  \subfigure[\label{fig:conv2} modified Gaussian]{
    \begin{minipage}{0.22\linewidth}
    \includegraphics[width=\textwidth]{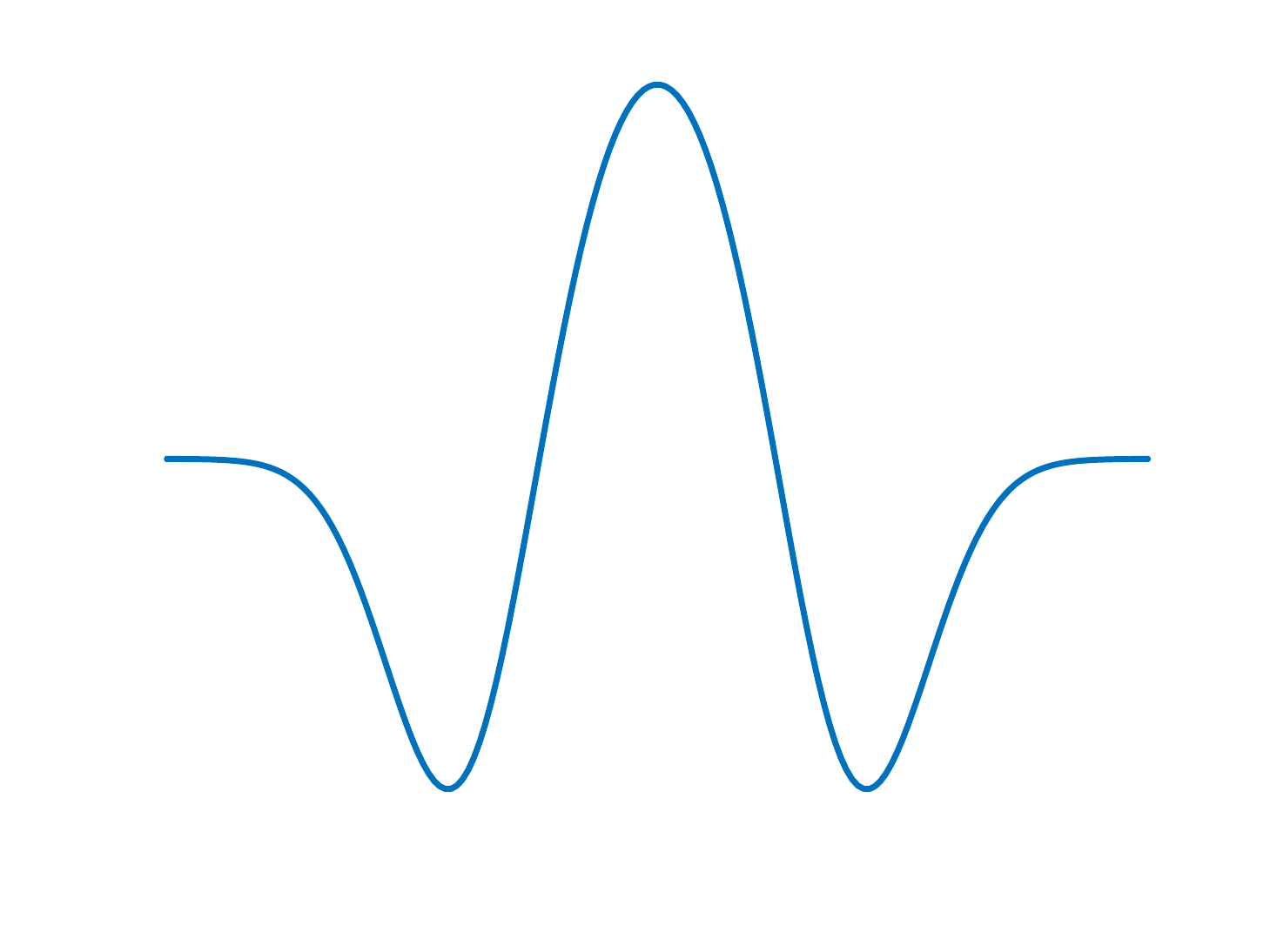}\\
    \includegraphics[width=\textwidth]{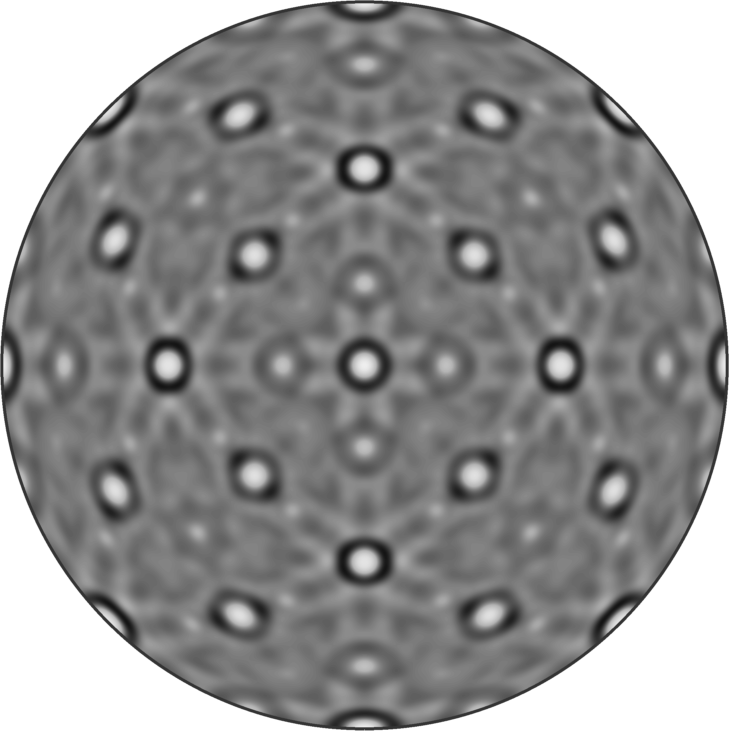}
    \end{minipage}
  }

  \caption{Spherical convolution of the master pattern with different band
    profiles.}
  \label{fig:conv}
\end{figure}

In Fig.~\ref{fig:conv211} and \ref{fig:conv310} the spherical convolutions of
the master pattern with band profiles corresponding to planes (211) and (310)
have been plotted. We observe extremely \changed{bright and sharp} peaks at
the corresponding band positions. The other bands are not as pronounced, as
they match the convolution template less well.

In order find a template $\Psi$ that reasonably fits all major bands we start
with a Gaussian profile in Fig.~\ref{fig:conv1} and modify it to
\begin{equation}
  \label{eq:profile}
  \Psi(\cos \omega) = \exp\Bigl(-\frac{(\omega-90)^{2}}{9}\Bigr)
  - \exp\Bigl(-\frac{(\omega-93)^{2}}{4}\Bigr)
   -\exp\Bigl(-\frac{(\omega-87)^{2}}{4}\Bigr)
\end{equation}
in \ref{fig:conv2} which we will rely on in our subsequent analysis.

Lets close this section by the remark that the convolution $f \star \Phi$ can
be computed as fast as the spherical Radon transform in Fourier space by the
formula
\begin{equation}
  \label{eq:RadonDelta}
  f \star \Phi(\vec \eta)
  = \sum_{n=0}^{N} \sum_{k=-n}^{n} \hat \Phi(n) \hat f(n,k) Y_{n}^{k}(\vec \eta)
\end{equation}
where $\hat \Phi(n)$ denotes the Legendre coefficients of the band profile
$\Psi$.

\subsection{Peak detection}
\label{sec:peak-detection-1}

In the conventional Radon/Hough transform approach for band detection in Kikuchi
pattern\changed{s}, the number of pixels in the Radon transform is approximately the
same as in the input (resized) Kikuchi pattern. This limits the possible
resolution of the orientations determined\footnote{The resolution of Radon/Hough based approaches is a combination of the resolution of the Radon space, the quality of the diffraction patterns, and the number of bands successfully localised and indexed.}. In contrast, when computing the
spherical Radon transform according to \eqref{eq:RadonS2} or the spherical
convolution by \eqref{eq:RadonDelta} such a restriction to a grid of pixels
does not exist. Instead, we can evaluate those sums for any normal vector
$\vec \eta$. Since it would be way too time consuming to evaluate
\eqref{eq:RadonS2} or \eqref{eq:RadonDelta} at an arbitrarily fine grid, we
propose a simultaneous steepest descent approach to find all local maxima.

The algorithm to find all peaks of a spherical function $g$ is as follows: we
start with a set of approximately equispaced points $\vec \eta_{m}$,
$m=1,\ldots,M$ on the sphere. Then we compute all the gradients
$\nabla g(\vec \eta_{m})$, $m=1,\ldots,M$ of $g$ according to the formulae
\begin{equation}
  \label{eq:grad}
  \begin{split}
    \partial_{\rho} g(\eta(\theta,\rho))
    = \sum_{n=0}^{N} \sum_{k=-n}^{n} \mathrm i k \hat g(n,k) Y_{n}^{k}(\vec \eta),\\
    \partial_{\theta} g(\eta(\theta,\rho))
    = \sum_{n=0}^{N} \sum_{k=-n}^{n} k \hat g(n,k) Y_{n}^{k}(\vec \eta),\\
    \nabla g(\eta)
    = \frac{1}{\sin \theta} \partial_{\rho} g(\vec \eta) \vec e_{\rho}(\vec \eta)
    + \frac{1}{\sin^2 \theta} \partial_{\theta} g(\vec \eta) \vec e_{\theta}(\vec \eta)
  \end{split}
\end{equation}
using the fast spherical Fourier transform and maximize $g$ locally along the
lines
\begin{equation*}
  \vec \eta_{m}^{1} = \vec \eta_{m} + \lambda_{m}^{1} \nabla g(\vec \eta_{m}),
  \quad \lambda_{m}^{1} \in [0,\pi).
\end{equation*}
This procedure is iterated and nodes $\vec \eta_{m}^{\ell}$ are found which
converge for $k \to \infty$ to the local maxima of the function $g$. During
the convergence, several of the nodes $\vec \eta_{m}^{\ell}$ will converge to
the same maxima and, hence, can be merged into one node.

\begin{figure}
  \centering
   \subfigure[\label{fig:bandsSphere} spherical projection]{
    \includegraphics[width = 0.45\textwidth]{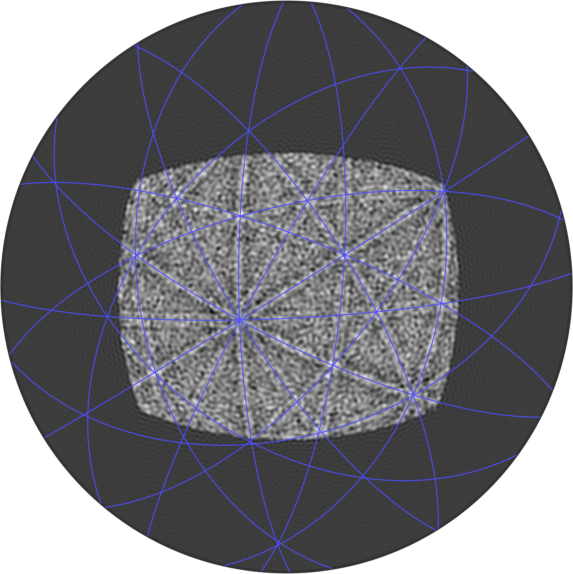}}
   \subfigure[\label{fig:peakDetection} spherical convolution]{
    \includegraphics[width = 0.45\textwidth]{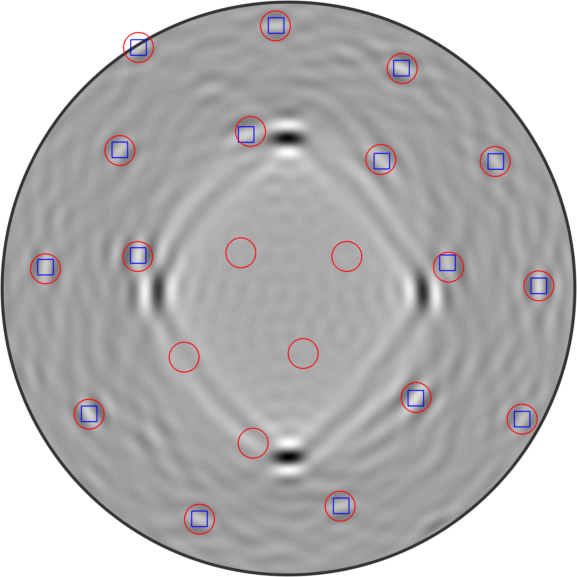}}
  \caption{(a) Simulated, noisy Kikuchi pattern (b) spherical convolution with
    the band profile from Fig.~\ref{fig:conv2}. The red circles mark the
    theoretical normals of the major lattice bands and the blue square the
    detected peaks. The corresponding bands are marked by blue circles in left
    picture.}
  \label{fig:IndexPat}
\end{figure}

\changed{To illustrate this procedure we apply it to the simulated, noisy
  Kikuchi pattern in Fig.~\ref{fig:bandsSphere} from which we computed the
  convolution with the modified Gaussian profile (Fig.~\ref{fig:conv}d) as
  depicted in Fig. \ref{fig:peakDetection}. The diamond shaped artifact in the
  center corresponds no normal vectors of the bands that are completely
  outside the detector region. The four vertices of the diamond correspond the
  the edges of the detector region. Since, the detector region is known we
  could adapt our peak finding algorithm to ignore all peaks inside the
  diamond as well as its vertices. Due to the noise in the Kikuchi pattern the
  convolution with the modified Gaussian profile contains many minor peaks not
  associated to any lattice plane. Nevertheless, selecting simply the 16
  brightest peaks (blue squares) found by our peak detection algorithm gave a
  very good coincidence with the theoretic positions of the major band normals
  (red circles). Obviously, the five red circles close to the diamond shaped
  artifact could not detected at all.}

\subsection{Orientation determination.}

Once a certain number of bands has been detected in the Kikuchi pattern any
of the well known indexing algorithms, e.g.~\cite{Morawiec15,Britton2018}, can
be used for determining the corresponding orientation. In this section we
utilise AstroEBSD \cite{Britton2018} and analyse the error distribution of the
resulting orientation determination method.

To refresh, our spherical Radon transform based \changed{orientation determination} method involves:

\begin{enumerate}
\item Project the experimental diffraction pattern onto the surface of the sphere
\item Approximate the discrete pattern by a spherical Fourier series as per equation
  \eqref{eq:S2Sum}.
\item Compute the spherical convolution \eqref{eq:RadonDelta} with a suitable
  band profile in Fourier space.
\item Detect the most pronounced peaks in the spherical convolution and search for their centres.
\item Use an indexing method (e.g.~AstroEBSD) to index bands and ultimately determine the crystal orientation.
\item If needed, this crystal orientation can be transformed into another frame (e.g.~from the detector frame to the sample frame).
\end{enumerate}

This algorithms involves a couple of parameters which need to be adjusted
carefully.  In step 1, the pattern centre must be known. In step 2, we have to
choose the harmonic cut-off frequency $N$, in step 3, a suitable band profile
(which matches the bands expected in our lookup table), and in step 4, the
number of iterations, the resolution of the initial search grid as well as the
number bands which will passed in step 5 into the indexing algorithm.

For our BCC-iron patterns we select the modified Gaussian band profile
(Fig.~\ref{fig:conv}d), calibrated using our master pattern, and set the
numbers of bands to 10.

\begin{figure}
    \begin{tikzpicture}
      \begin{axis}[width=0.55\textwidth,height=6cm,
        ymin=0,
        xmin=0,
        xmax=1,
        xlabel={misorientation angle in degree},
        legend style={font=\footnotesize},
        xlabel style={font=\footnotesize},
        ybar interval,
        xtick=,
        xticklabel={$\pgfmathprintnumber\tick - \pgfmathprintnumber\nexttick$}
        ]
        \addplot+ [hist={data=x,bins=20}] file {sim/band_48_14.txt};
        \addplot+ [hist={data=x,bins=20}] file {sim/band_64_14.txt};
        \addplot+ [hist={data=x,bins=20}] file {sim/band_96_14.txt};
        \legend{$N=48$,$N=64$,$N=96$}
      \end{axis}
    \end{tikzpicture}
    \begin{tikzpicture}
      \begin{axis}[width=0.55\textwidth,height=6cm,
        ymin=0,
        xmin=0,
        xmax=1,
        xlabel={misorientation angle in degree},
        legend style={font=\footnotesize},
        xlabel style={font=\footnotesize},
        ybar interval,
         xtick=,
        xticklabel={$\pgfmathprintnumber\tick - \pgfmathprintnumber\nexttick$}
        ]
        \addplot+ [hist={data=x,bins=20}] file {sim/band_64_10.txt};
        \addplot+ [hist={data=x,bins=20}] file {sim/band_64_12.txt};
        \addplot+ [hist={data=x,bins=20}] file {sim/band_64_14.txt};
        \legend{$b=10$,$b=12$,$b=14$}
      \end{axis}
    \end{tikzpicture}

    \caption{Histograms of the misorientation between the original orientation
      and the orientation determined by spherical band detection. Left
      histogram fixes the
      number of bands $b=14$ and varies the harmonic cut-off degrees $N$ and
      right histogram fixes the harmonic cut-off degrees $N=64$ and varies the
      number of bands $b$.}
  \label{fig:histBand}
\end{figure}
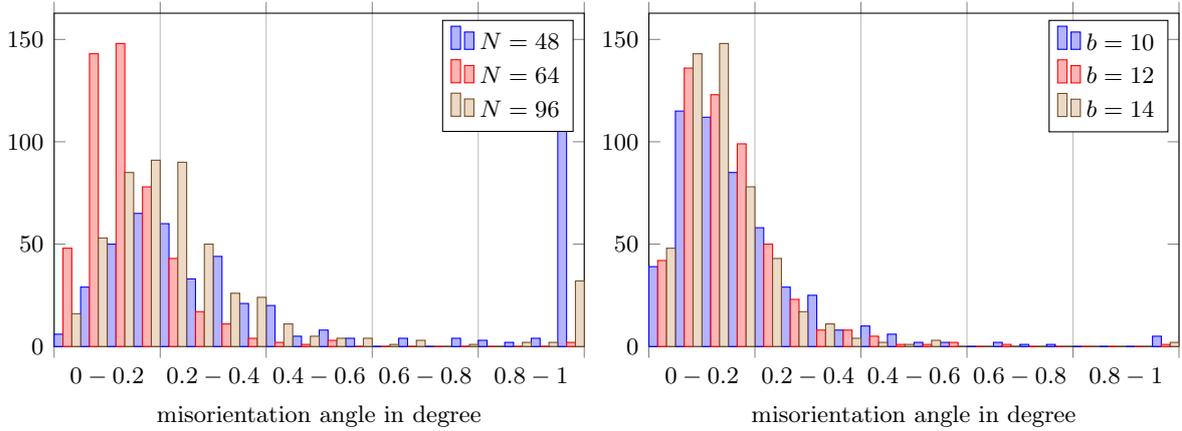

Putting everything together we first verify our method with simulated noisy
patterns. Therefore, we proceed as follows. First we select a random orientation
$\vec O$. Then we dynamically simulate a corresponding Kikuchi pattern with
$400 \times 300$ pixel and add noise as displayed in
Fig.~\ref{fig:bandsSphere}.  We use this pattern to determine an orientation
$\vec{\tilde O}$. Finally, we compute the misorientation between initial
orientation $\vec O$ and the computed orientation $\vec{\tilde O}$. Histograms
of these misorientation angles for different harmonic cut-off degrees $N$ are
depicted in Fig.~\ref{fig:histBand}. We determine that a mean accuracy of
$0.1\degree$ can be obtained \changed{when the pattern centre is known exactly \textit{a priori}}.

In Sec.~\ref{sec:exper-demonstr} we will demonstrate this orientation
determination method with an experimental data set.

\section{Spherical cross correlation based \changed{orientation determination}}
\label{sec:SphereXCFindex}
We have established that experimental and master pattern can be well
represented by their harmonic expansion on the sphere and that this
representation is useful for band detection. Now we present the use of this
representation when computing the cross correlation between an experimental
pattern with all possible rotations of a master pattern.

Template matching of EBSD patterns usually employs the following steps:
\begin{enumerate}
\item Simulate a dynamical master pattern of all orientation vectors.
\item Select a dense set of orientations $\zb O_{m}$, $m=1,\ldots,M$.
\item \changed{Create a library of Kikuchi patterns $S^{m}_{ij}$ with respect
    to all orientations $\zb O_{m}$, $m=1,\ldots,M$ by rotating and projecting
    the master to the detector plane.}
\item For a measured experimental pattern $P_{ij}$ compute the cross
  correlations $C(m)$, $m=1,\ldots,M$ with respect to all patterns $S^{m}_{ij}$
  within the library.
\item Select the orientation $\zb O_{\tilde m}$ with the largest cross correlation
  value $C(\tilde m)$ as the indexed orientation.
\end{enumerate}
The main advantage of this template matching based \changed{approach for
  orientation determination} is that it takes into account all diffraction
pattern features and does not reduce the analysis to a simple ``geometry''
based problem of localising and indexing the bands. This provides the
potential that this method is more robust to noise.

The main disadvantage of the template matching approach is that reprojection
of the master pattern for a dense population of orientation space is memory
intensive and that the repeated computation of the cross correlation of each
experimental pattern with all patterns of the library is computationally expensive.

To overcome this shortcoming, we transfer the template matching based approach to
our spherical setting and make use of the fast Fourier transform on the
rotation group. This allows us to compute the spherical cross correlations
simultaneously for all orientations $\zb O_{m}$, $m=1,\ldots,M$ which is much
faster than by a pixel by pixel based formula.

\subsection{Spherical cross correlation}
\label{sec:spher-cross-corr}

We start by representing both the simulated pattern as well as the
experimental pattern by expansions in spherical harmonics
\begin{align}
  f_{\text{sim}}(\zb h)
  &= \sum_{n=0}^{N} \sum_{k=-n}^{n} \hat f_{\text{sim}}(n,k)
  Y_{n}^{k}(\zb h)\\
  \label{eq:2}
  f_{\text{exp}}(\zb r)
  &= \sum_{n=0}^{N} \sum_{k'=-n}^{n} \hat f_{\text{exp}}(n,k')
  Y_{n}^{k'}(\zb r)
\end{align}
as discussed in Sec.~\ref{SphereFT}. Note, that the simulated pattern
$f_{\text{sim}}$ is usually represented with respect to crystal coordinates,
while the experimental pattern $f_{\text{exp}}$ is represented with respect to
detector coordinates. Let $\zb O$ be the exact crystal orientation of the
experimental pattern, i.e., $\zb r = \zb O \zb h$. Then the basic assumption
of the pattern matching approach is that the simulated pattern transformed
into the specimen reference frame gives a good approximation of the
experimental pattern modulo a scaling factor $\alpha \in \mathbb R$, i.e.
\begin{equation*}
  f_{\text{exp}}(\zb r) \approx   \alpha f_{\text{sim}}(\changed{\zb O^{-1} \zb r}).
\end{equation*}

The similarity of two spherical functions modulo a rotation $\zb O$ can be
measured by the spherical cross correlation, which is defined as the integral
of the product of both functions over the entire sphere
\begin{equation}
  \label{eq:S2Cor}
  C(f_{\text{sim}},f_{\text{exp}})(\zb O)
  = \int_{\mathbb S^{2}} f_{\text{sim}}(\zb O^{-1} \zb r) f_{\text{exp}}(\zb r)
  \ \mathrm d\zb r
  = \int_{\mathbb S^{2}} f_{\text{sim}}(\zb h) f_{\text{exp}}(\zb O \zb h)
  \ \mathrm d\zb h.
\end{equation}

In order to evaluate these integrals numerically one could make use of a
spherical quadrature rule with nodes $\zb h_{n} \in \mathbb S^{2}$ and weights
$\omega_{n} > 0$, $n=1,\ldots,N$, cf. Sec.~\ref{sec:quadrature}, which leads
to the sum
\begin{equation}
  \label{eq:S2CorQuad}
  C(f_{\text{sim}},f_{\text{exp}})(\zb O)
  \approx \sum_{n=1}^{N} \omega_{n}  f_{\text{sim}}(\zb h_{n}) f_{\text{exp}}(\zb O \zb h_{n}).
\end{equation}
This sum does not require to pre-compute and store
a dictionary of simulated patterns. Instead it is sufficient to store the
simulated master pattern $f_{\text{sim}}$ at the quadrature nodes $\zb h_{n}$ and
transform each experimental pattern to the sphere. Furthermore, it can
be evaluated at arbitrary orientations $\zb O$, i.e, we are not restricted to
any grid in the orientation space.

\subsection{Fast evaluation of the spherical cross correlation}
\label{sec:fast-eval-spher}

A critical disadvantage of the template matching approach are its high
computational costs. Indeed, evaluating the cross correlation function
\eqref{eq:pm} at a dense set of $M$ orientations for patterns with $N^{2}$
points requires $M \cdot N^{2}$ numerical operations. Evaluating the spherical
cross correlation function \eqref{eq:S2CorQuad} directly would lead to the
same amount of numerical operations. In this section we show how fast Fourier
techniques on the orientation space can be exploited to speed up this
computation to only $N^{3} \log N + M$ numerical operations.

The idea is to use the following important relationship between spherical
harmonics $Y_{n}^{k}$ and Wigner-D functions $D_{n}^{k,k'}$,
cf. \cite{HielscherPotts2008},
\begin{equation*}
  D_{n}^{k,k'}(\zb O)
  = \int_{\mathbb S^{2}} Y_{n}^{k}(\zb O \zb r)
  \overline{Y_{n}^{k'}(\zb r)} \ \mathrm d \zb r
\end{equation*}
which allows us to rewrite the series expansion of the rotated simulated
pattern as
\begin{equation}
  \label{eq:1}
    f_{\text{sim}}(\zb O^{-1} \zb r)
    = \sum_{n=0}^{N} \sum_{k=-n}^{n} \sum_{k'=-n}^{n} D_{n}^{k,k'}(\vec O)
    \hat f_{\text{sim}}(n,k') Y_{n}^{k'}(\zb r).
\end{equation}
Inserting the series expansions \eqref{eq:2} and \eqref{eq:1} into the
correlation integral \eqref{eq:S2Cor} and making use of the orthogonality of
the spherical harmonic $Y_{n}^{k}$ we end up with
\begin{align}
  C(f_{\text{sim}},f_{\text{exp}})(\zb O)
  &= \sum_{n=0}^{N} \sum_{k,k' = -n}^{n}
  \hat f_{\text{sim}}(n,k) \hat f_{\text{exp}}(n,k')
  D_{n}^{k,k'}(\zb O)\\
  &=\sum_{n=0}^{N} \sum_{k,k' = -n}^{n}
    \hat C(n,k,k') D_{n}^{k,k'}(\zb O),
    \label{eq:convSO3}
\end{align}
where we have set
$\hat C(n,k,k') = \hat f_{\text{sim}}(n,k) \hat f_{\text{exp}}(n,k')$. The
latter sum \eqref{eq:convSO3} is known as the Fourier transform on the
rotation group and can be evaluated at $M$ arbitrary orientations $\zb O_{m}$
using only $N^{3} \log N + M$ numerical operations by the algorithm described
in \cite{Potts2009}.

\begin{figure}
  \centering
  \subfigure[]{
  \includegraphics[height=0.4\textwidth]{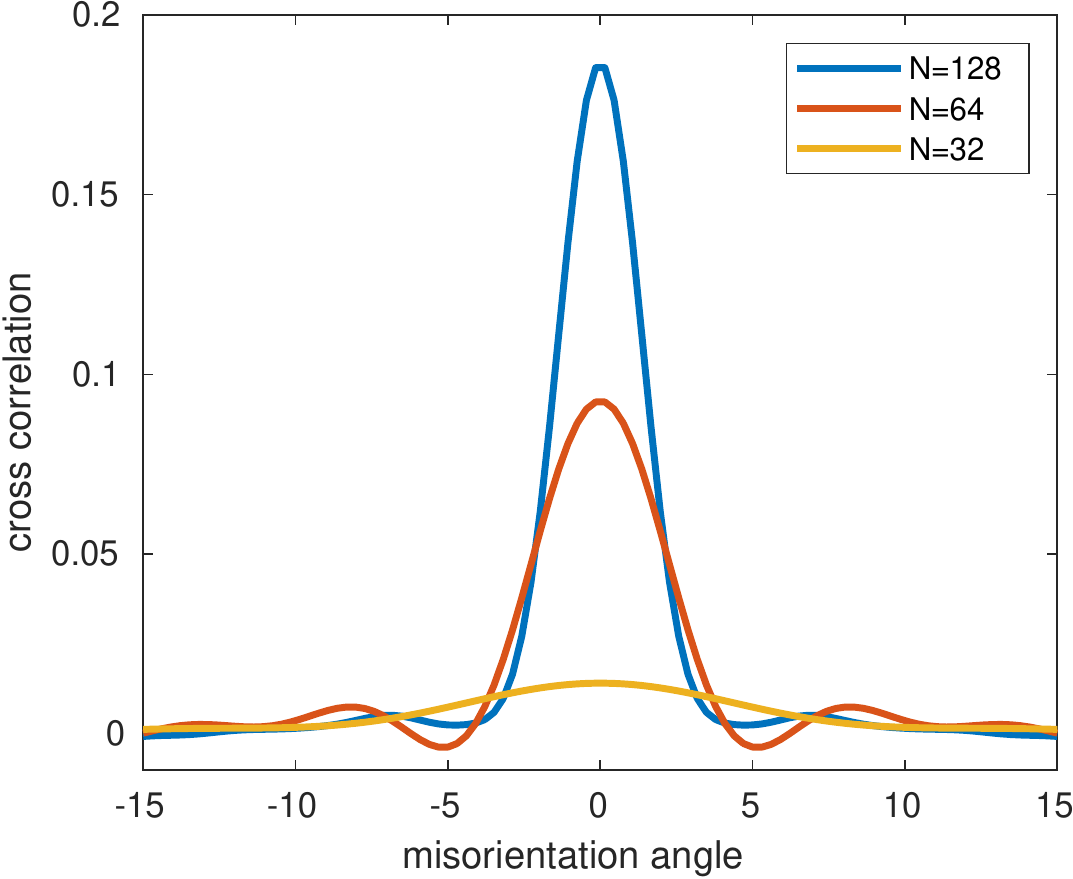}}
  \subfigure[]{
    \includegraphics[height=0.4\textwidth]{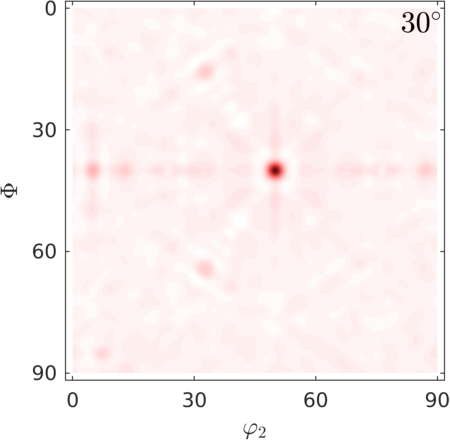}}
\centering
  \subfigure[\label{fig:xcorPhi1Pattern}]{
    \includegraphics[height=0.4\textwidth]{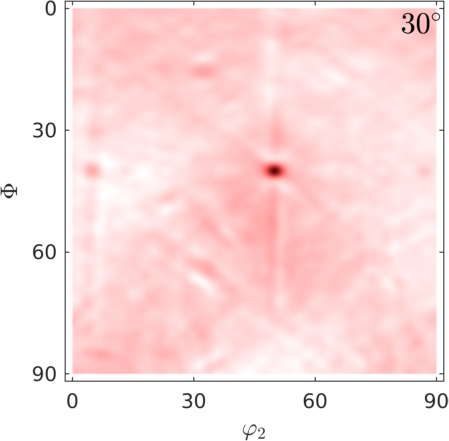}}
  \subfigure[\label{fig:xcorCorrected}]{
  \includegraphics[height=0.4\textwidth]{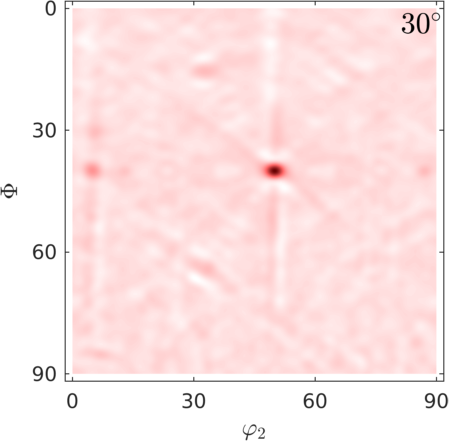}}
\caption{Spherical cross correlation: (a) as a function of the misorientation
  angle from the exact orientation with respect to different harmonic cut-off
  parameters $N$; (b) as a section through the Euler space showing the
  dominant peak (dark red) at the position of the exact match; (c) the same
  section but with an ``experimental'' pattern, showing the artifacts due to
  the incomplete coverage of the sphere by the detector window; (d) the
  corrected cross correlation function $C(\vec O)$ according to \eqref{eq:3}.}
  \label{fig:xcorSigma}
\end{figure}

In order to illustrate our approach we have chosen a random orientation
$\zb O$ and defined the function
$f_{\text{exp}}(\zb r) = f_{\text{sim}}(\zb O^{-1} \zb r)$ as a rotated version
of our master pattern. In a second step we approximated both pattern\changed{s} by
expansions into spherical harmonics up to cut-off degree $N=512$. Finally, we calculated the spherical cross correlation function
$C(f_{\text{sim}},f_{\text{exp}})$ as a function of the misorientation
from the initial orientation $\zb O$ for different, smaller, cut-off degrees $N$. The
results are depicted in Fig.~\ref{fig:xcorSigma}a. We observe that a cut
off degree $N=64$ gives a good localisation of the peak position close to the true
orientation.

\subsection{Correction}
\label{sec:correction}

In the previous section we have assumed that the test pattern $f_{\text{exp}}$
is known at the entire sphere. In practice, however, only the projection of
the detector back to the sphere is known. This causes low frequency artifacts
in the cross correlation function as depicted in Fig.~\ref{fig:xcorSigma}c.

Luckily, these artefacts can be computed explicitly as the spherical cross
correlation $C(f_{\text{sim}},\chi)$ between the simulated Kikuchi pattern $f_{\text{sim}}$ on
the sphere and the cut-off function $\chi$ of the detector region projected to the
sphere. The final difference
\begin{equation}
  \label{eq:3}
  C(\zb O) = C(f_{\text{sim}},f_{\text{exp}})(\zb O) - \frac{\int_{\mathbb S^{2}}
    f_{\text{exp}}}{\int_{\mathbb S^{2}} \chi}  C(f_{\text{sim}},\chi)(\zb O)
\end{equation}
is depicted in \ref{fig:xcorCorrected}.

\subsection{Peak detection}
\label{sec:peak-detection}

Peak detection for functions of the form \eqref{eq:convSO3} can be implemented
in a similar manner as explained in Sec.~\ref{sec:peak-detection-1} for
spherical functions since the gradient can again be written as a sum with
respect to Wigner-D functions. However, in the present case we are only
interested in finding the global maximum (and not many local maxima). This
makes it efficient to evaluate the corrected cross correlation function
\eqref{eq:3} on a fixed  and uniformly spaced grid of orientations $\zb O_{m}$,
$m=1,\ldots,M$ with resolution $\delta^{(1)} \approx 2\degree$ and choose the
orientation $\zb O_{\tilde m}$ with maximum function value $C(\zb O_{\tilde m})$. In a
second step we choose a local grid around the orientation $\zb O_{\tilde m}$ with
radius $\delta^{(1)}$ and resolution $\delta^{(2)} \approx 0.1\degree$ and
repeat the calculation. The global resolution needs to be chosen such that no
peak falls between the grid points.

\subsection{Accuracy determination for spherical cross correlation}

We perform a numerical experiment to test the accuracy of our spherical cross
correlation algorithm and optimise crucial parameters such as the harmonic cut\changed{-}off degree $N$ as well as the resolutions $\delta^{(1)}$ and $\delta^{(2)}$ of
the global and local search grids:
\begin{enumerate}
\item Compute a spherical Fourier series approximation $f_{\text{sim}}$ of a
  dynamically simulated master pattern as described in
  Sec.~\ref{sec:interpolation}.
\item Generate random crystal orientation $\zb O$ and simulate a corresponding
  ``experimental Kikuchi pattern'' by projecting the master pattern to a virtual
  detector and adding Poisson distributed noise.
\item Project the noisy ``experimental pattern'' back to the sphere, multiply it by the mask
  $\varphi$ and approximate the product by a spherical Fourier series $f_{\text{exp}}$.
\item Evaluate the corrected spherical cross correlation function
  $C(\zb O_{m})$ between $f_{\text{sim}}$ and $f_{\text{exp}}$ at the grid
  orientations $\zb O_{m}$.
\item Determine the grid orientation $\zb O_{\tilde m}$ with the largest cross
  correlation value.
\item Compute the misorientation angle between the initial random orientation $\zb O$
  and the computed orientation $\zb O_{\tilde m}$.
\end{enumerate}

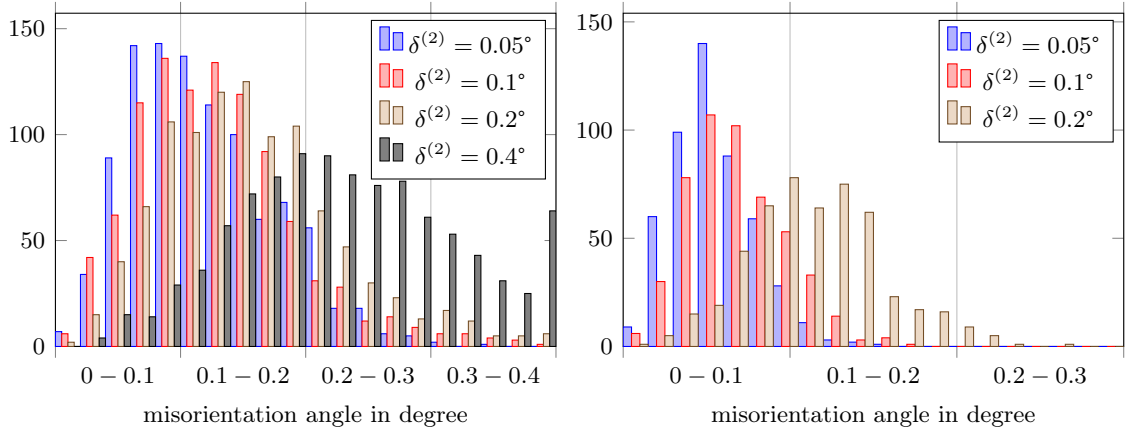
\begin{figure}
  \centering

    \begin{tikzpicture}
      \begin{axis}[
        height=6cm,width=0.525\textwidth,
        ymin=0,
        xmin=0,
        xmax=0.4,
        xlabel={misorientation angle in degree},
        ybar interval,
        xlabel style={font=\footnotesize},
        legend style={font=\footnotesize},
        xtick=,
        xticklabel={$\pgfmathprintnumber\tick - \pgfmathprintnumber\nexttick$}
        ]
        \addplot+ [hist={data=x,bins=20}] file {sim/s2xcor_48_125_05.txt};
        \addplot+ [hist={data=x,bins=20}] file {sim/s2xcor_48_125_10.txt};
        \addplot+ [hist={data=x,bins=20}] file {sim/s2xcor_48_250_20.txt};
        \addplot+ [hist={data=x,bins=20}] file {sim/s2xcor_48_500_40.txt};
        \legend{$\delta^{(2)}=0.05\degree$,$\delta^{(2)}=0.1\degree$,
          $\delta^{(2)}=0.2\degree$,$\delta^{(2)}=0.4\degree$}
      \end{axis}
    \end{tikzpicture}
    \begin{tikzpicture}
      \begin{axis}[
        height=6cm,width=0.525\textwidth,
        ymin=0,
        xmin=0,
        xmax=0.3,
        xlabel={misorientation angle in degree},
        ybar interval,
        xlabel style={font=\footnotesize},
        legend style={font=\footnotesize},
        xtick={0,0.1,0.2,0.3},
        xticklabel={$\pgfmathprintnumber\tick - \pgfmathprintnumber\nexttick$}
        ]
        \addplot+ [hist={data=x,bins=20}] file {sim/s2xcor_64_150_05.txt};
        \addplot+ [hist={data=x,bins=20}] file {sim/s2xcor_64_150_10.txt};
        \addplot+ [hist={data=x,bins=20}] file {sim/s2xcor_64_250_20.txt};
        \legend{$\delta^{(2)}=0.05\degree$,$\delta^{(2)}=0.1\degree$,$\delta^{(2)}=0.2\degree$}
      \end{axis}
    \end{tikzpicture}

  \caption{Misorientation angle histograms between the ``true'' random
    orientation used for simulating the diffraction pattern and the
    orientation determined by spherical cross correlation. Left histogram
    fixes harmonic cut-off bandwidth $N=48$ and right histogram $N=64$. Only
    the resolution $\delta^{(2)}$ of the refined grid is given. The
    corresponding resolution of the global grid can be found in Table \ref{tab:s2xcor}.}
\label{fig:s2xcorHist}
\end{figure}

This numerical experiment has been run 500 times for different choices of the
harmonic cut-off bandwidth\changed{s} $N$ and \removed{the} different resolutions of the search
grid $\zb O_{m}$. Table \ref{tab:s2xcor} summarises the parameters, the run
times and the achieved precision. Full histograms of the misorientation angles
are depicted in Figure \ref{fig:s2xcorHist}.

We observe that a global resolution $\delta^{(1)} = 5\degree$ is to\changed{o} coarse as it
leads to about 5 percent completely mis-indexed patterns. For all other
parameter choices we obtain reasonable angular precision up to $0.05\degree$
with a speed of one pattern per second on an ordinary laptop without any
graphic card support.


%

\begin{table}
  \centering
  \begin{tabular}{rrrrrrrrr}
    \toprule
    \multicolumn{1}{c}{cut-off}
    & \multicolumn{2}{c}{global search grid}
    & \multicolumn{3}{c}{local search grid}
    & speed
    & \multicolumn{2}{c}{precision} \\
    $N$
    & res. $\delta^{(1)}$
    & points $M_{1}$
    & radius
    & res. $\delta^{(2)}$
    & points $M_{2}$
    & pattern/s
    & median
    & std \\
    \midrule
      48 & $5\degree$ & $4\,958$ & $5\degree$ & $0.4\degree$
    & $9\,106$ & 1.8 &0.22&4.56\\
    48 & $2.5\degree$ & $39\,565$ & $2.5\degree$ & $0.2\degree$
    & $9\,128$ & 1.5 &0.15&0.07\\
    48 & $1.5\degree$ & $183\,035$ & $1.5\degree$ & $0.1\degree$
    &  $14\,005$ & 1.3 &0.12&0.06\\
    48 & $1.5\degree$ & $183\,035$ & $1.5\degree$ & $0.05\degree$
    & $112\,514$ &1.1 &0.11&0.06\\
    64 & $2.5\degree$ & $39\,565$ & $2.5\degree$ & $0.2\degree$
    & $9\,128$ & 1.4 &0.11&0.04\\
    64 & $1.5\degree$ & $183\,035$ & $1.5\degree$ & $0.1\degree$
    &  $14\,005$ & 1.2 &0.06&0.03\\
    64 & $1.5\degree$ & $183\,035$ & $1.5\degree$ & $0.05\degree$
    & $112\,514$ &1.0 &0.05&0.02\\
    \bottomrule
  \end{tabular}
  \caption{Indication of computational costs and associated precision for
    spherical cross correlation. Times are measured on an ordinary laptop.}
  \label{tab:s2xcor}
\end{table}

\section{Experimental demonstration}
\label{sec:exper-demonstr}

\begin{figure}
  \centering

  \begin{tabular}{ccc}
    & inverse pole figure (IPF) colouring
    & axis angle colouring \\
    \raisebox{0.9\height}{\rotatebox[origin=l]{90}{2D Radon}}
    & \includegraphics[width=0.45\textwidth]{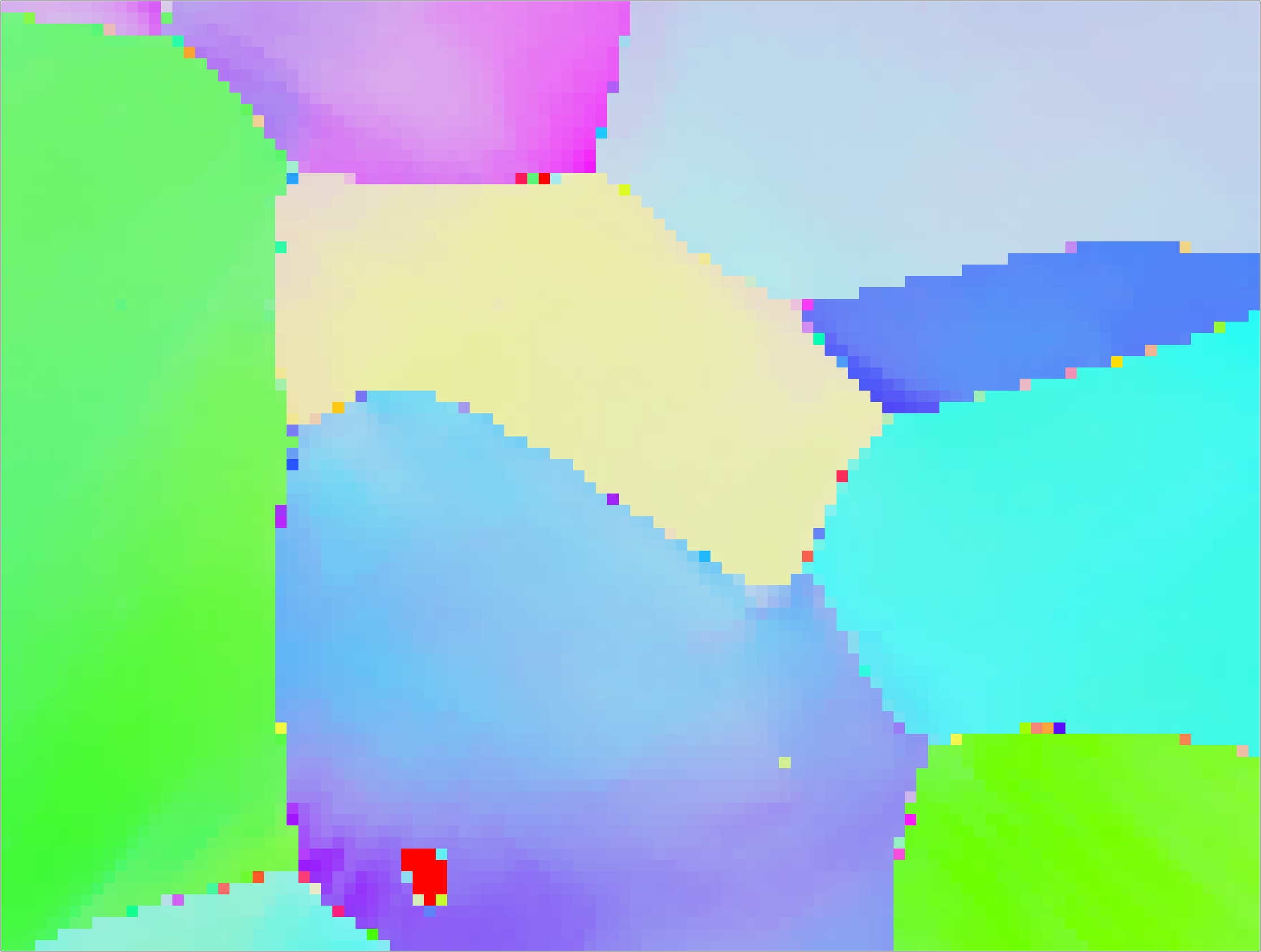}
    & \includegraphics[width=0.45\textwidth]{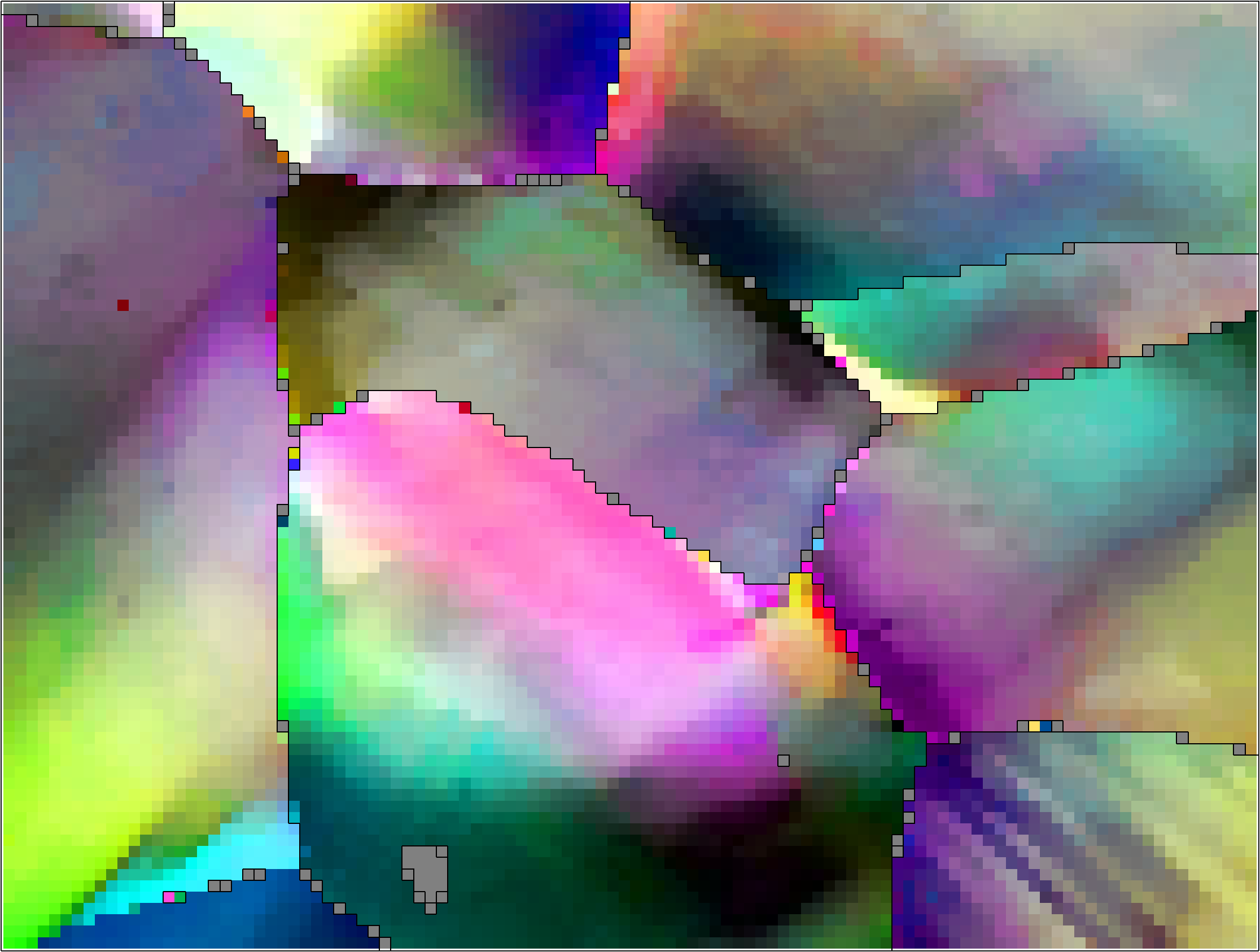}   \\
    \raisebox{0.3\height}{\rotatebox[origin=l]{90}{spherical Radon}}
    & \includegraphics[width=0.45\textwidth]{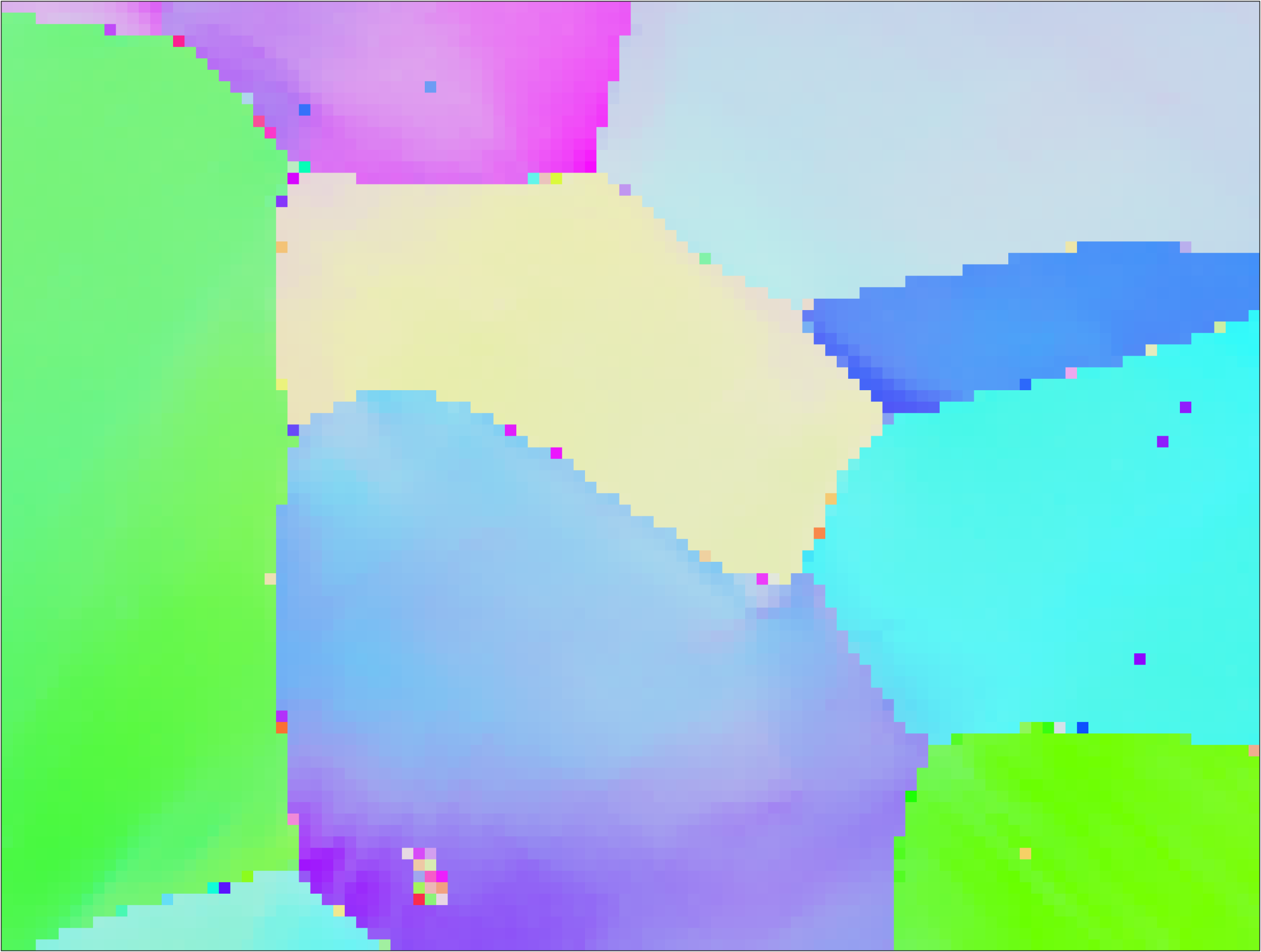}
    & \includegraphics[width=0.45\textwidth]{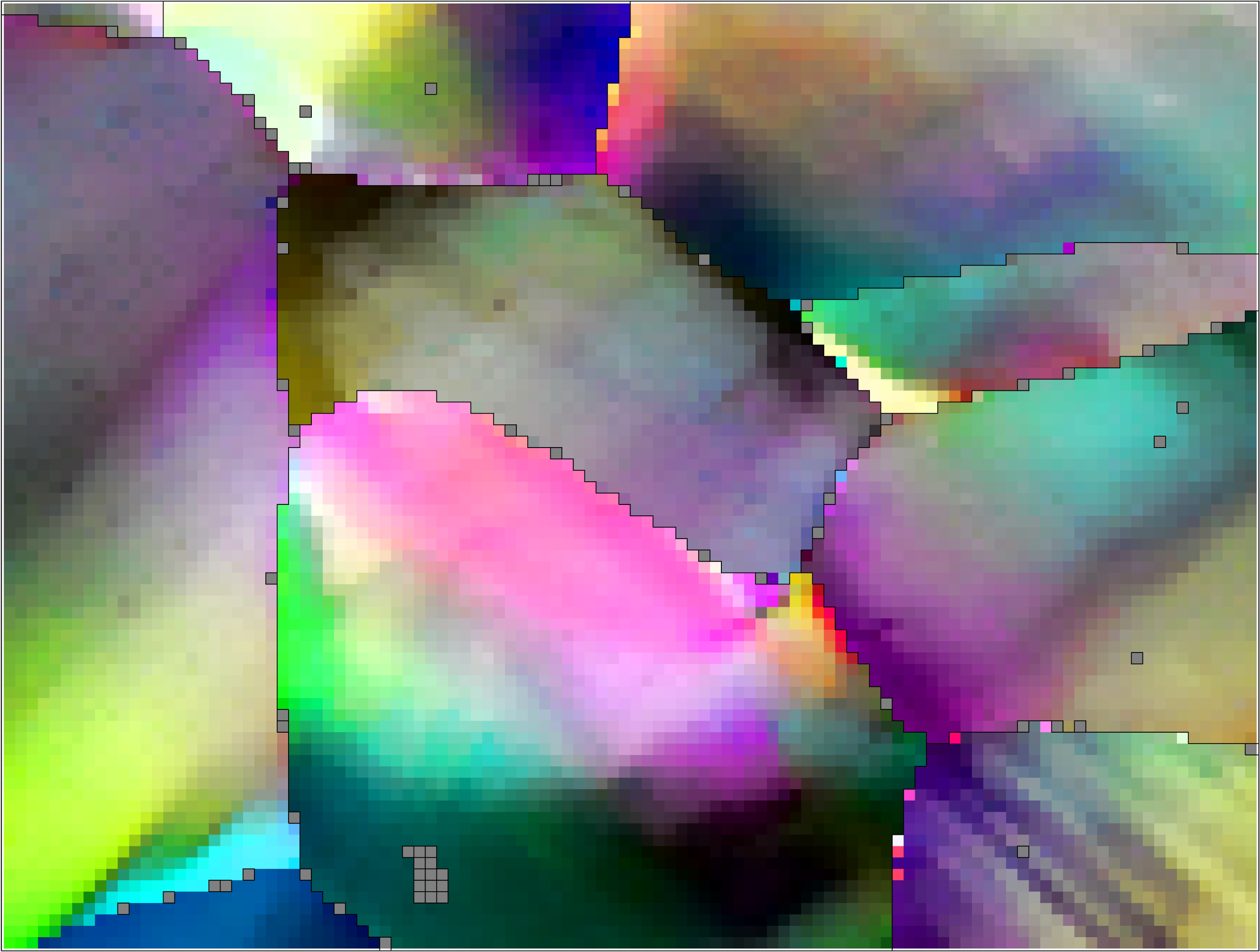} \\
    \raisebox{+0.1\height}{\rotatebox[origin=l]{90}{spherical cross correlation}}
    & \includegraphics[width=0.45\textwidth]{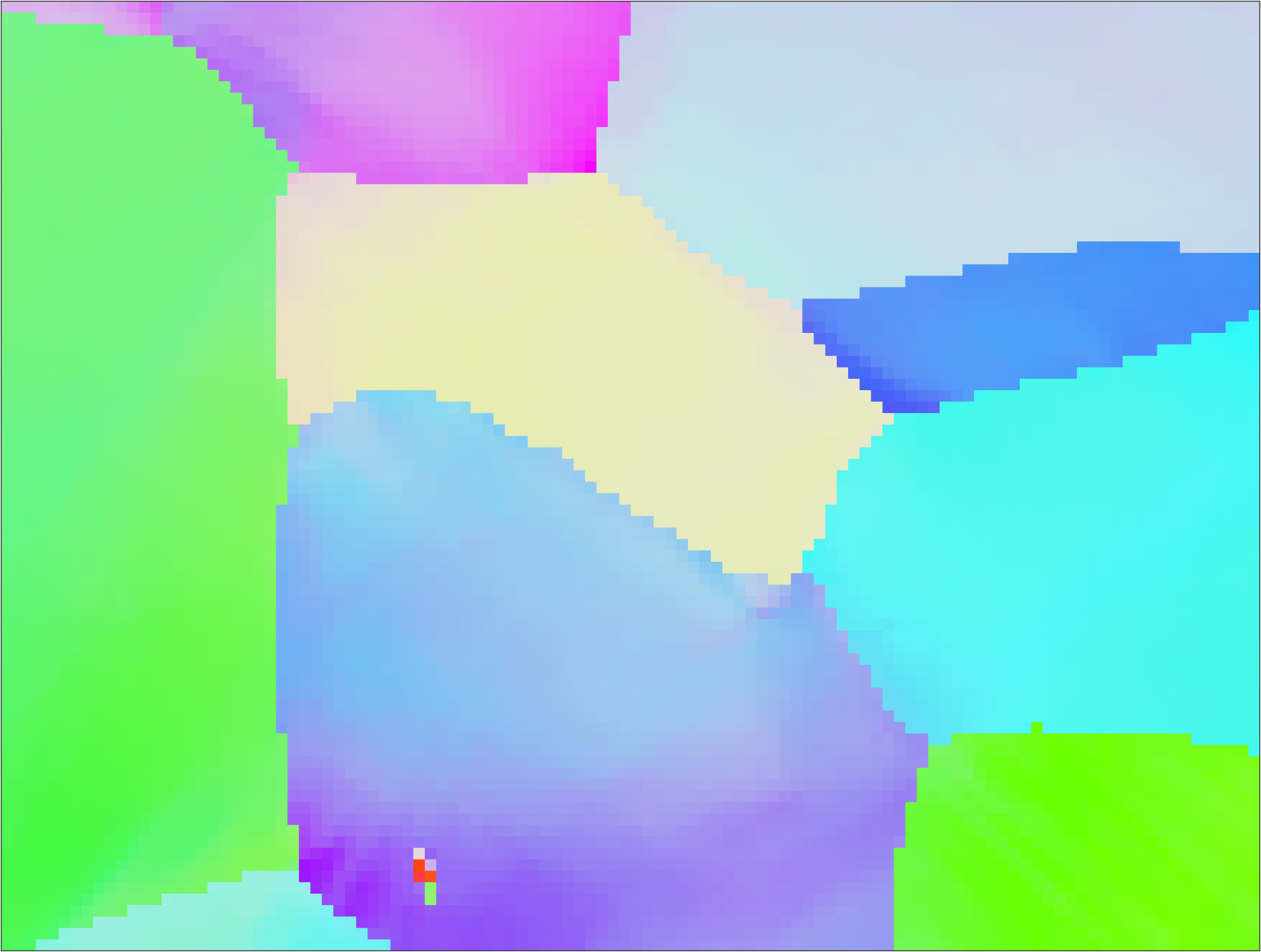}
    & \includegraphics[width=0.45\textwidth]{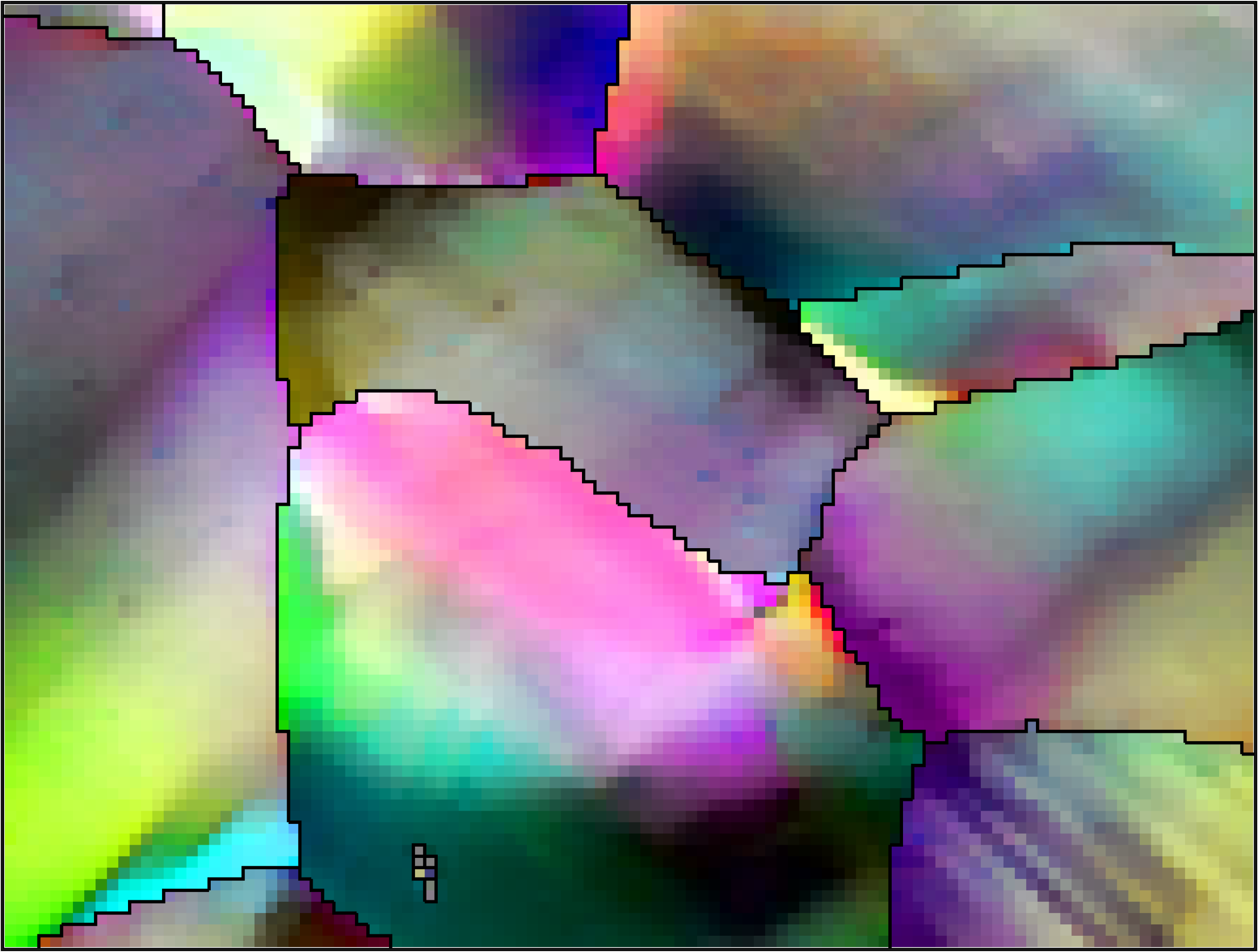} \\
    \raisebox{0.3\height}{\rotatebox[origin=l]{90}{color keys}}
    & \includegraphics[width=0.15\textwidth]{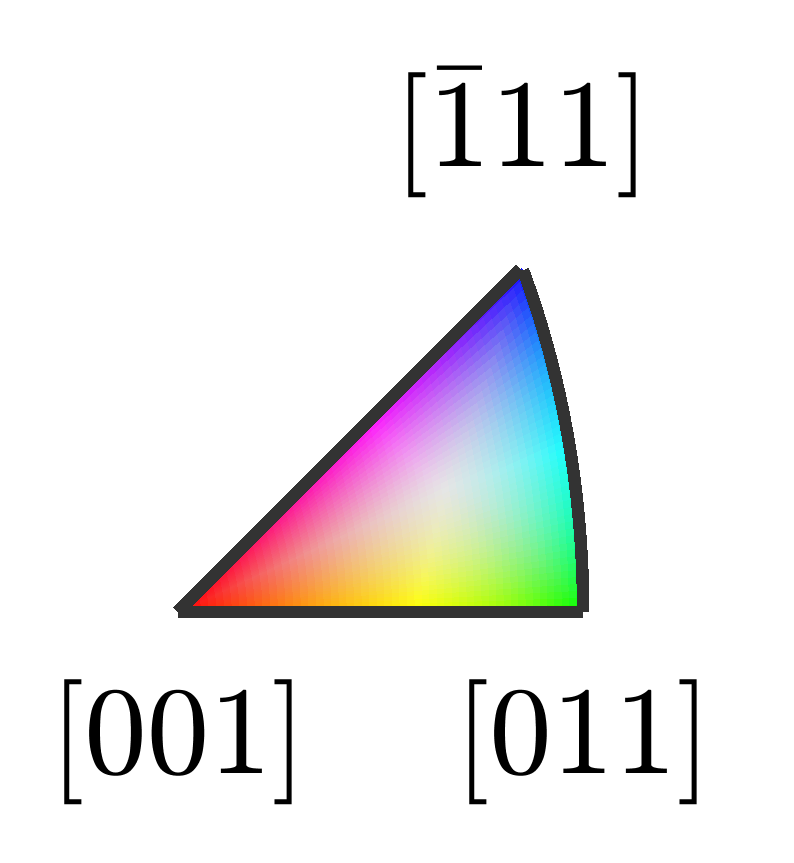}
    & \includegraphics[width=0.3\textwidth]{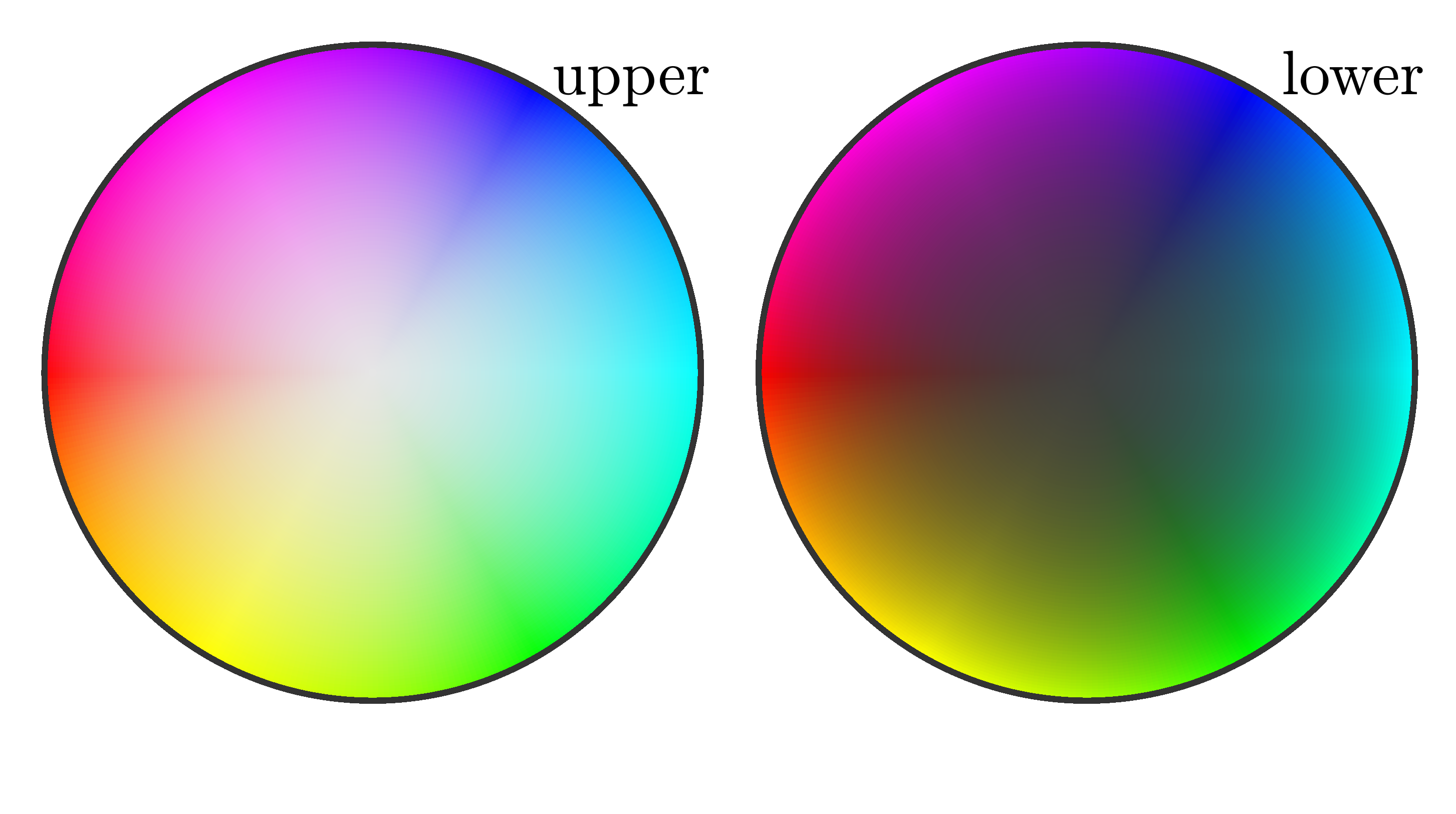}   \\
  \end{tabular}
  \caption{Demonstration $\alpha$-iron data indexed using 2D based Radon transform band localisation and AstroEBSD indexing, spherical Radon transform band localisation and AstroEBSD indexing, and spherical cross correlation. The IPF colour key is with respect to the horizontal axis. The axis-angle colour key is taken with respect to the mean grain orientation and each key has a radius of $5\degree$.}
  \label{fig:orig}
\end{figure}

We test our two spherical algorithms using the demonstration $\alpha$-Iron data
set as used in Britton et al.~\cite{Britton2018} for conventional indexing
using a planar Radon transform and the AstroEBSD indexing algorithm. This data
can be found on Zenodo \url{https://zenodo.org/record/1214829} and consists of
a 9\,130 point EBSD pattern map. The AstroEBSD background correction was used
with operations: hot pixel correction; resize to 300 pixels wide; low
frequency Gaussian background division (sigma = 4), performed independently on
each detector half, circular radius cropping to 0.95 of the pattern width.
All peak ID based indexing was performed using the iron phase file, with the
top 10 bands used in the analysis.  The flat Radon transform based analysis
was performed with 1 degree theta resolution and up to 13 peaks were
sought. The pattern centre was optimised by searching for the minimum weighted
mean angular error using a $10 \times 10$ grid array.  The spherical Radon
transform based orientation determination was performed using the idealised
profile given in \eqref{eq:profile} and the spherical cross correlation based
\changed{orientation determination} was performed using the harmonic
cut-of\changed{f} frequency $N=64$ and resolutions
$\delta^{(1)} = 1.5\degree$, $\delta^{(2)} = 0.1\degree$ for the global,
respective, local search grid. Results are presented in
Fig. \ref{fig:orig}. The orientations are by all three method reasonably
recovered. The smoothness in the axis-angle plots are similar, where the
spherical method performs slightly better near grain boundaries and the spherical
cross correlation method is significantly more robust near grain boundaries.

\section{Discussion}

In this manuscript, we demonstrated that considering EBSD patterns as
spherical images allows for elegant algorithms for band detection, band
analysis and cross correlation. The reason behind this elegance is the fact
that in its spherical representation symmetry operations and misorientations
are simple rotations of the Kikuchi pattern.

Beside being elegant these algorithms can be implemented using fast Fourier
techniques which makes them at least theoretically fast. In practice, our
algorithms do not yet meet the speed of highly optimized implementations of
standard Hough transform based indexing algorithms. The reason for this is
that our algorithms are not yet parallelized, do not take advantage of
symmetries and are mainly implemented with readability in mind.

Based on spherical band detection and spherical cross correlation we presented
two new methods for orientation determination from Kikuchi pattern. In
numerical experiments with noisy, simulated as well as experimental patterns we
achieved an accuracy of up to $0.1$ degree. Clearly, this accuracy depends on
the noise level and the resolution of the provided EBSD pattern. A more
precise relationship between noise level and resolution on the one side and
achievable accuracy on the other side is subject of further research. Our
examples on the iron data set demonstrated that our algorithm is more robust
then conventional flat Radon transform based approaches.

Another advantage of the spherical analysis is that spherical convolution can
used to extract specific bands according to their profile. Furthermore, those
profiles can be efficiently computed from experimental patterns using formula
\eqref{eq:4} and then be explored with respect to shape and symmetry. This has
the potential to better understand the asymmetry created either by an improper
pattern centre \cite{Bassinger2011}, a subtle changes in the lattice
\cite{Maurice2008}, and band asymmetry \cite{Winkelmann2016,Winkelmann2008}.



An important assumption of the spherical method is that we know the pattern
centre \textit{a priori}. The pattern centre is important for our analysis, as
an incorrect pattern centre will introduce a distortion on the rendering of
patterns on the sphere and the band edges would no longer be
parallel. Computationally it can be expensive to optimise the pattern centre
based upon this constraint, but there have been suggestions in the literature
that centre on this idea (e.g.~the 3D Hough \cite{Maurice2008} and one of the
methods of the BYU group \cite{Bassinger2011} in attempting absolute strain
measurement with high resolution EBSD).  If we assume that we know, or can
measure, the pattern centre with reasonable precision with standard methods
(e.g.~just using the 2D Radon based pattern centre measurements available
within commercial or open-source software such as AstroEBSD
\cite{Britton2018}), then we can perform re-projection and gain reasonable
indexing success which is demonstrated with the example iron data set shown
with this work. Perhaps more excitingly, it is likely that the spherical
approach will prove useful when less conventional capture geometries are used,
as the divergence of the Kossel cones is naturally encoded when the pattern is
projected onto the sphere. For instance, it is well known that this divergence
has caused significant problems when analysing transmission Kikuchi patterns
when the pattern centre no longer located within the detector screen.

\section{Conclusion}
We have outlined and demonstrated methods to perform analysis of EBSD patterns in a spherical frame. We can summarise our conclusions as:
\begin{itemize}
\item  Simulated as well as experimental Kikuchi patterns can be well
  approximated by spherical functions. These approximations can be computed
  efficiently using the fast spherical Fourier transform.

\item The choice of a suitable harmonic cut-off frequency is crucial for the approximation
  process.

\item The spherical Radon transform and spherical convolution are efficient
  methods for band detection in Kikuchi pattern.

\item The spherical approximation allows for an efficient method for
  extracting band profiles.

\item The spherical cross correlation is an efficient method for
  determining the orientation of a Kikuchi pattern by comparing it with a
  rotated versions of a master pattern.

\item The spherical Radon transform, spherical convolutions, as well as
  spherical cross correlation can be efficiently computed using fast Fourier
  transforms on the sphere and the rotation group.

\item Spherical approximation, the spherical Radon transform, spherical
  convolution and spherical cross correlation can be adapted to work well with
  patterns that do not cover the entire sphere as it is typical for
  experimental pattern measured at a flat detector.

\item In our numerical experiments with simulated, noisy Kikuchi patterns the
  spherical Radon transform based methods as well as the spherical cross
  correlation based methods for orientation determination achieved a precision
  of $<0.1\degree$.

\end{itemize}

\section{Data statement}
The example iron data set can be found on Zenodo (https://doi.org/10.5281/zenodo.1214828). Upon article acceptance the full code for this manuscript will be released to Zenodo.

\section*{Acknowledgements}
TBB acknowledges funding of his research fellowship from the Royal Academy of
Engineering. We thank Alex Foden for useful discussions regarding pattern
matching. We thank Jim Hickey for assisting with the example iron data set
which was captured in the Harvey Flower EM suite at Imperial College on
equipment supported by the Shell AIMS UTC. We thank Aimo Winkelmann for
assistance with the spherical reprojection and dynamical pattern generation.
\changed{Finally, we thank the anonymous reviewers for their valuable comments and suggestions.}

\bibliography{references}

\end{document}